\documentclass[preprint2]{aastex631}
\NewPageAfterKeywords

\usepackage{amsmath}
\usepackage{amssymb}
\usepackage{amsfonts}
\newcommand{\ssun}{S_{\earth}}
\newtheorem{hypothesis}{Hypothesis}
\usepackage{chemformula}
\usepackage{mathtools}
\usepackage{float}

\begin{document}

\title{Interior convection regime, host star luminosity, and predicted atmospheric \ch{CO2} abundance in terrestrial exoplanets}

\correspondingauthor{Antonin Affholder}
\email{antoaffh@gmail.com}

\author[0000-0003-3481-0952]{Antonin Affholder}
\affiliation{Department of Ecology and Evolutionary Biology, University of Arizona, Tucson AZ, USA}

\author{Stéphane Mazevet}
\affiliation{Observatoire de la Côte d'Azur, Université Côte d'Azur, Nice, France}

\author[0000-0001-6164-756X]{Boris Sauterey}
\affiliation{Département de Géosciences, ENS, PSL, Paris, France.}

\author[0000-0003-3714-5855]{D\'aniel Apai}
\affiliation{Steward Observatory, University of Arizona, Tucson AZ, USA}
\affiliation{Lunar and Planetary Laboratory, University of Arizona, Tucson AZ, USA}
             
\author[0000-0002-5806-5566]{Régis Ferrière}
\affiliation{Department of Ecology and Evolutionary Biology, University of Arizona, Tucson AZ, USA}
\affiliation{Institut de Biologie de l'École Normale Supérieure, ENS, PSL, Paris, France}
\affiliation{International Research Laboratory for Interdisciplinary Global Environmental Studies (iGLOBES), CNRS, ENS, PSL, University of Arizona, Tucson AZ, USA}



\begin{abstract}

Terrestrial planets in the Habitable Zone of Sun-like stars are priority targets for detection and observation by the next generation of space telescopes.
Earth's long-term habitability may have been tied to the geological carbon cycle, a process critically facilitated by plate tectonics.
In the modern Earth, plate motion corresponds to a mantle convection regime called mobile-lid.
The alternate, stagnant-lid regime is found on Mars and Venus, which may have lacked strong enough weathering feedbacks to sustain surface liquid water over geological timescales if initially present.
Constraining observational strategies able to infer the most common regime in terrestrial exoplanets requires quantitative predictions of the atmospheric composition of planets in either regime.
We use endmember models of volcanic outgassing and crust weathering for the stagnant- and mobile-lid convection regimes, that we couple to models of atmospheric chemistry and climate, and ocean chemistry to simulate the atmospheric evolution of these worlds in the Habitable Zone.
In our simulations under the two alternate regimes, we find that the fraction of planets possessing climates consistent with surface liquid water is virtually the same.
Despite this unexpected similarity, we predict that a mission capable of detecting atmospheric \ch{CO2} abundance above 0.1 bar in 25 terrestrial exoplanets is extremely likely ($\geq 95$\% of samples) to infer the dominant interior convection regime in that sample with strong evidence (10:1 odds).
This offers guidance for the specifications of the Habitable Worlds Observatory NASA concept mission and other future missions capable of probing samples of habitable exoplanets.

\end{abstract}



\section{Introduction} \label{sec:intro}

The question of whether Earth's habitability is generic or specific among rocky planets in the Habitable Zone \citep[HZ, the range of orbital radii as a function of stellar mass for which rocky planets with surface liquid water do not undergo greenhouse nor global glaciation runaway;][]{huang1959occurrence,Hart1979,kasting1993habitable,kopparapu2013habitable} is central to the search for extraterrestrial life.
Planets of composition and size similar to the Earth's --terrestrial exoplanets-- in the HZ of Sun-like (FGK-type stars) have not yet been observed and thus are priority targets for detection and characterization by the next generation of ground-based and space telescopes \citep{NAP26141}.
Exoplanet atmospheric compositions inferred from such observations may serve to confirm or reject that these planets are habitable in the sense that liquid water is actually present on their surface \citep{Robinson2018,cockell2016habitability}.\\

\noindent
These analyses, however, are likely to come with large degrees of uncertainty and ambiguity. 
To alleviate such ambiguity, three steps can be taken.
First, atmospheric exoplanet data should be interpreted given the planet's context by taking into account planetary characteristics such as orbit radius, host star mass, planet radius, mass, age, and formation pathways \citep{meadows2018exoplanet,Apai2018,bixel2020testing,krissansen2022understanding}.
Second, Bayesian inference makes it possible to combine contextual information, used to shape prior distributions, with observations, to derive posterior probabilities of habitability and inhabitation \citep{catling2018exoplanet}.
Finally, the assessment of habitability and putative inhabitation will be more robust by analyzing patterns of atmospheric composition across a sample of exoplanets, rather than the detailed characterization of any single planet \citep{Bean2017,bixel2021bioverse,mazevet2023prospects}.
In recent years, there has been some pioneering effort to implement this agenda \citep[\textit{e.g.} ][]{lehmer2020carbonate,bixel2020testing}.
Yet we are still lacking predictions of population-level atmospheric patterns of habitability and potential for life emergence under alternate geophysical hypotheses, and a quantitative evaluation of their testability.\\

\noindent
The HZ framework typically provides limit values of luminosity above and below which no surface liquid water is expected \citep{kopparapu2013habitable,kasting1993habitable}.
However, these analyses cannot predict how likely the presence of surface liquid water is on planets within the HZ boundaries.
Constraining the climate of the Earth in the distant geological past is difficult.
\citet{krissansen2017constraining} used a parametric model of the geological carbon cycle \citep{Kasting1993Earth} and climate to show that the evolution of atmospheric \ch{CO2} as estimated from fossil weathering profiles \citep{Rye1995,Hessler2004,Sheldon2006,Kanzaki2015} is consistent with the persistence of a relatively temperate climate as the Sun became increasingly bright.
Drawing a parallel between the changing luminosity of the Sun and varying orbital radii around a star of constant brightness, it is commonly hypothesized that geophysical processes assumed to have stabilized the Earth's climate --the geological carbon cycle-- would ensure that planets within the HZ have climates compatible with surface liquid water \citep{kasting1993habitable,catlingdavid2018exoplanet}.
In other words, the concept of the HZ as currently defined rests largely on the hypothesis of the geological carbon cycle as a global negative feedback loop to climate.\\

\noindent
\citet{lehmer2020carbonate} proposed testing the hypothesis that an Earth-like geological carbon cycle acts as a planetary thermostat that ensures temperate climates across the range of incident stellar light flux in the HZ.
Their model of the Earth's geological carbon cycle predicts a correlation between luminosity and atmospheric p\ch{CO2}.
In their simulations, sampling about 83 exoplanets is required to reject the null hypothesis of a log-uniform (uncorrelated) distribution, a number far greater than the target yield of the NASA Flagship concept Habitable Worlds Observatory \citep{mamajek2023}.\\

\begin{figure}
    \centering
    \includegraphics[scale=1]{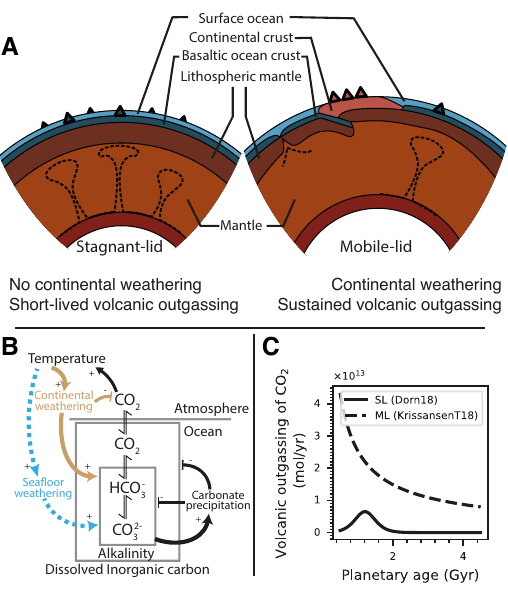}
    \caption{\textbf{Key geophysical and geochemical processes of stagnant-lid (SL) and mobile-lid (ML) convection regimes in the interior-ocean-atmosphere coupled model.} \textbf{A} Key geophysical processes. The dashed curves in the mantle represent isotherms highlighting convective plumes. \textbf{B} Key geochemical processes and their influence of global temperature. Brown and blue arrows indicate the contribution of continental (brown arrows) and seafloor (blue arrows) to the climate-stabilizing negative feedback. Triangle-headed arrows indicate positive effects (increase in quantity), whereas flat-headed arrows indicate negative effects (removal or decrease). Two-ways arrows represent dissolution or chemical equilibrium. Brown arrows correspond to processes exclusive to the silicate-carbonate feedback. Dotted blue arrows correspond to processes exclusive to the basalt dissolution (seafloor weathering) feedback. \textbf{C} \ch{CO2} volcanic outgassing rate as a function of planetary age as proposed for the Earth, that we use as the ML regime nominal model \citep[dashed curve; KrissansenT18 indicates model from][]{krissansen2018constraining} and for one Earth mass SL planets \citep[continuous curve; Dorn18 indicates model from][]{dorn2018outgassing}}
    \label{fig:schema}
\end{figure}

\noindent
While this work provides an important first step, its calculation of the false negative decision risk (frequency of non rejection of the assumed false null hypothesis) relies on an \textit{ad hoc} null hypothesis with no explicit geophysical basis.
That Mars, Venus, and the early Earth likely lacked continental crust powering modern Earth-like geological carbon cycle provides an empirical null or alternate hypothesis to that of a negative correlation p\ch{CO2}-luminosity caused by both continental and seafloor weathering.
Even in the absence of a geological carbon cycle, photochemical reactions in the atmosphere may affect the abundance of \ch{CO2} over geological timescales.
In addition, continental weathering relies primarily on continental crust minerals (silicates and carbonates) being exposed to precipitations.
The accretion of continental crust is dependent upon plate tectonics, a process that our planetary neighbors Mars and Venus do not appear to have or have had \citep{solomatov1996stagnant,catling2017atmospheric}, and that may have had a late onset on Earth \citep[although this view is being challenged; see][]{Windley2021}.
No planet in our solar system appears to have been in any one clearly defined interior convection regime during the entirety of its geological history.
However, it appears that two distinct end-member scenarios that are relevant to a planet's habitability can be drawn:
That the potential habitability of terrestrial planets be supported by the geological carbon cycle as enabled by Earth-like plate tectonics equated with the "mobile-lid" (ML) interior convection regime; or as an alternate hypothesis that terrestrial planets in general remain in the "stagnant-lid" (SL; also referred to as single-lid) interior convection regime, in which no crust recycling or continental crust accretion occur (Figure~\ref{fig:schema}A).\\

\noindent
It is often assumed that for conditions favorable to surface habitability, in particular a temperate climate persisting on geological timescales, weathering of continental rocks (silicates and carbonates; 'continental weathering') is required \citep{catling2017atmospheric}.
This assumption, in combination with Earth's unique interior convection regime among terrestrial planets in the solar system, may lead to expect that most terrestrial planets orbiting in the habitable zone of other stars may not effectively possess surface liquid water.
However, weathering of non-continental emerged land may still participate to the geological carbon cycle, and the seafloor dissolution of basalt can also drive a negative feedback to atmospheric \ch{CO2} that does not hinge on the existence of continental crust  \citep[see Figure~\ref{fig:schema}B, and][]{krissansen2018constraining,gillis2011secular}.
Seafloor weathering may allow terrestrial planets with a limited supply of subaerial weatherable silicates and carbonates to retain a geological carbon cycle as a negative feedback loop to climate \citep{Lenardic2016}, albeit a possibly weaker one \citep{sleep2001}.
Planets in the SL regime are assumed to have a shorter-lived period of volcanic outgassing due to the absence of recycling of crust-bound \ch{CO2} and the thickening of the lithosphere as the mantle cools \citep[Figure~\ref{fig:schema}C;][]{dorn2018outgassing}.
Nevertheless, early on they may support an ocean and sufficient basalt production to sustain controlled atmospheric \ch{CO2} through seafloor weathering.
Thus, even though continental weathering conservatively remains the main process for long-term Earth-like habitability, the findings of first-order analyses of planets atmosphere in the SL regime call for models that fully couple interior, ocean, and atmospheric processes to evaluate the habitability of SL worlds \citep{foley2018carbon,foley2019habitability,Tosi2017}.\\

\noindent
How likely terrestrial planets in the HZ might be and remain habitable, or even possibly host life, could thus largely depend on the influence of a convection regime on climate, and on what convection regime is the most common.
Under the seemingly pessimistic assumption that terrestrial exoplanets in the HZ are in the SL regime, how likely are they to possess stable temperate climates? What are the key planetary parameters that this depends on?
And how can this inform the design of future exoplanet survey missions to assess how widespread SL and ML interior convection regimes are, at minimal cost?
To answer these questions, we present an integrated model in which atmospheric photochemistry is coupled with ocean chemistry, weathering processes, and climate.
We model equilibrium atmospheric composition in the two concurrent interior convection regime by adopting endmember models for relating volcanic outgassing to planetary age \citep{krissansen2018constraining,dorn2018outgassing}, and set the weathering of sub-aerial rocks to be zero in the SL regime.
Models for the evolution of outgassing rate on Earth-sized planets remain scarce and large uncertainties are associated with them, especially in the case of the stagnant-lid regime.
Given these uncertainties, we focus on a nominal model for Earth's history for the ML convection regime \citep{krissansen2018constraining}, and the model outputs from \citet{dorn2018outgassing} for the SL convection regime.
The coupled model offers a significant step towards the generalization of early-Earth models that include the biological activity of microbial biospheres \citep{sauterey2020}.
The model also extends previous work in which photochemistry as well as outgassed greenhouse gases other than \ch{CO2}, such as methane or dihydrogen, were omitted \citep{lehmer2020carbonate,foley2018carbon}.
By feeding simulations of our model into a Bayes factor based experimental design analysis, we predict which combination of minimally detectable \ch{CO2} and sample size offers the highest chance of inferring the dominant interior convection regime with sufficient evidence.

\section{Methods}

Our goal is to simulate the atmospheric composition and habitability of terrestrial exoplanets under various conditions of luminosity and for two regimes of interior convection: the stagnant-lid (SL) and the mobile-lid (ML) regimes.
To do so, we couple slightly modified versions of models describing the influence on atmospheric composition of ocean and atmospheric chemistry, as well as chemical exchanges between the atmosphere and the crust, the ocean and the atmosphere, and the crust and the ocean \citep{arney2016pale,krissansen2017constraining}.
Thus, we generalize existing coupled models beyond the particular case of the Archean Earth \citep{kharecha2005coupled,sauterey2020}.
In this section, we present the description of each of these models and their assemblage into our coupled model.
Then, simulations resulting from this coupled model are used to perform statistical analyses that project the potential yield of future observational strategies as described in Results Section \textit{Simulated inference of dominant interior convective regime among terrestrial exoplanets}.\\

\noindent
First, we describe how our coupled model is built.
We expand on models designed for simulating the climate of the Archean Earth \citep{sauterey2020,krissansen2018constraining,kharecha2005coupled}.
We select a restricted number of molecules of which the abundance in the atmosphere is modeled.
These are chosen for their importance in determining the planet's surface temperature (loosely referred to as climate throughout the text) and thus potential habitability, their participation to geochemical processes relevant to climate, and their potential use for an early biosphere similar to Earth's early biosphere \citep{sauterey2020}:\ch{H2}, \ch{CO2}, \ch{CH4}, and \ch{CO}.\\

\noindent
Several key planetary parameters could affect the equilibrium value of the abundance of those gases in the atmosphere of a terrestrial planet.
Volcanic outgassing ($F_{out}$, mol~yr$^{-1}$), taken broadly, releases most of those molecules (and others) into the atmosphere, has played a key role in setting the climate of the early Earth (and still does so today), and has changed over geological times on Earth \citep{krissansen2018constraining}.
It is also believed that volcanic outgassing may vary in its intensity depending on the interior convection regime a planet is exhibiting.
Second, the quantity of light (luminosity) a planet receives from its star is expected to be a first-order determining factor of its climate (its value relative to that of the modern Earth is noted $S/\ssun$).
Last, weathering cycles that participate to set atmospheric \ch{CO2} require that fresh weatherable minerals are exposed to the planetary surface, hence it depends on volcanic activity on the one hand, and on the surface of crust that is exposed to precipitations, denoted by the fraction of the planetary surface $f_{land}$.
Together with other parameterization terms of the geological carbon cycle model (Table \ref{table:initialization}), these compose the parameter vector $\boldsymbol{\theta}$.
Other than for those parameters, it is assumed that the simulated planet is identical to the Earth (one Earth mass, 1 bar atmosphere).\\

\noindent
Ultimately, we aim at building a system of differential equations that describes the rate of change of the vector composed of partial pressures in the atmosphere $\boldsymbol{y}=(p\ch{H2}, p\ch{CO2}, p\ch{CH4}, p\ch{CO})$ as a function $\mathrm{F}$ of $\boldsymbol{y}$ itself and the parameters $\boldsymbol{\theta}$:
\begin{equation}
\label{eq:der_gen}
    \dot{\boldsymbol{y}}=\mathrm{F}_{\boldsymbol{\theta}}(\boldsymbol{y})
    ~.\\
\end{equation}

\noindent
If the geological timescale (on which $\boldsymbol{\theta}$ changes) is much slower than the climate timescale (on which $\boldsymbol{y}$ changes), the system of differential equations in Equation (\ref{eq:der_gen}) is seen as \emph{autonomous}, i.e. the function $\mathrm{F}$ does not explicitly depend on time $t$.
In the following sections, we develop the components of the function $\mathrm{F}_{\boldsymbol{\theta}}(\boldsymbol{y})$ that involve atmospheric photochemistry (Methods Section \textit{Atmospheric photochemistry and climate}), exchanges between the ocean and the atmosphere (Methods Section \textit{Molecular flux across the ocean/atmosphere interface}), and ocean chemistry of carbonates and carbon monoxide including its relation to geochemical processes such as weathering (Methods Section \textit{Ocean-Atmosphere coupling}).

\subsection{Atmospheric photochemistry and climate}
Our goal is to derive a numerical method to obtain the rate of change of atmospheric mixing ratios due to a set photochemical reactions denoted $\boldsymbol{\Phi}$ (molecules~cm$^{-2}$~s$^{-1}$), the surface temperature $T_s$ (K), and the planetary albedo ($A_P$) given the atmospheric mixing ratios $\boldsymbol{y}$ (in Pa):
\begin{equation}
    \label{eq:atmeq}
    (\boldsymbol{\Phi},T_s,A_P) = \mathrm{F}^{atm}_{\boldsymbol{\theta}}(\boldsymbol{y})~.\\
\end{equation}

\noindent
Studies that aim to determine the boundaries of the Habitable Zone (HZ) adopt an inverse modeling approach where outgoing longwave radiation is calculated for various temperatures \citep{kasting1993habitable,kopparapu2013habitable}.
However, we aim to roughly calculate climatic conditions and atmospheric composition of terrestrial planets that lie \emph{within} the HZ by assumption.
This introduces a number of differences in the way atmospheric modeling is handled in our coupled model, which are detailed in this section.\\

\noindent
We use \textit{Atmos}, a 1-D coupled photochemistry and climate model \citep{arney2016pale}, that we run over ranges of values for (p\ch{H2}, p\ch{CO2}, p\ch{CH4},$S/\ssun$).
Then, the output simulations are interpolated to create a numerical function which approximates $\mathrm{F}_{\boldsymbol{\theta}}^{atm}$ in Equation (\ref{eq:atmeq}).
The \textit{Atmos} model simulates the equilibrium vertical profile (with respect to photochemistry and climate) of the abundance of molecules in the atmosphere given a set of boundary conditions that are either (i) fixed molecular fluxes or (ii) fixed gas mixing ratios at the bottom-most layer of the discretized atmosphere.
Using boundary conditions defined as fixed deposition or outgassing rates instead of fixed molecular abundances would lead the model to calculate the vertical composition of the atmosphere at the equilibrium between the imposed surface fluxes and photochemical reactions.
This would then lead to a discrepancy between the state of the atmosphere calculated by the photochemical model and the out-of-equilibrium atmospheric composition that is used to calculate geochemical fluxes (weathering and ocean/atmosphere exchanges) that constitutes the set of dynamic variables of the model $\boldsymbol{y}$ (Equation \ref{eq:der_gen}).
In order to use this model to obtain photochemical rates when the lower atmosphere has a composition given by vector $\boldsymbol{y}$, we need to run it using defining boundary conditions as fixed mixing ratios $\boldsymbol{y}=(p\ch{H2}, p\ch{CO2}, p\ch{CH4}, p\ch{CO})$.\\

\begin{figure*}
    \centering
    \includegraphics[scale=1]{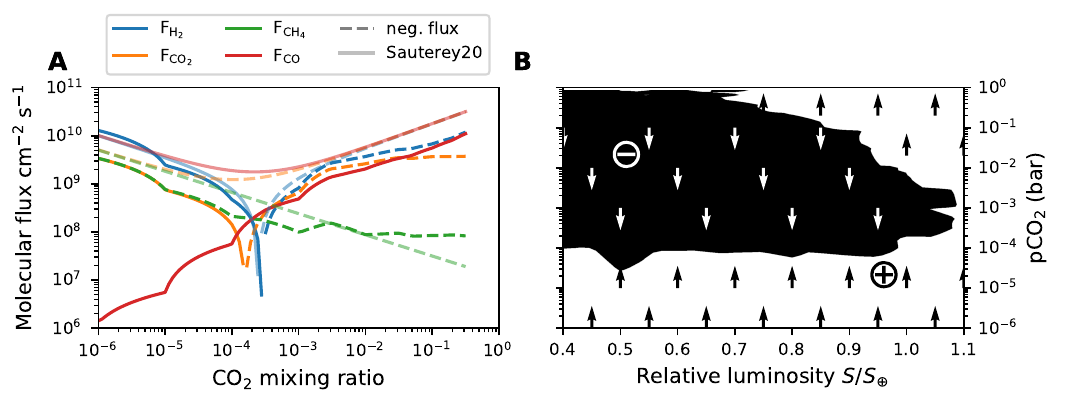}
    \caption{\textbf{A} Photochemical rates of production (solid lines) and alteration (dashed lines) of molecules included in our coupled model calculated from our interpolation of \textit{Atmos} outputs or calculated from the parameterization in \citet{sauterey2020} (semi-transparent lines).
    These calculations assume $S/\ssun=0.8$, the value of the luminosity assumed for the Archean Earth in the parameterization of \citet{sauterey2020}.
    Other gases are fixed at p$\ch{CH4}=0.1$~ppm and p$\ch{H2}=100$~ppm, average values in our simulations.
    \textbf{B} Sign of the photochemical flux of \ch{CO2} as a function of atmospheric concentration in \ch{CO2} and relative luminosity.
    Dark areas represent p\ch{CO2}-luminosity conditions in which the photochemistry results in net removal of \ch{CO2}, and the white region of the plot represent conditions in which photochemistry results in net production of \ch{CO2}.
    These calculations use fixed p$\ch{CH4}=0.01$~ppm and p$\ch{H2}=35$~ppm, representative of SL simulations with age $>2$~Gyrs which have a high final \ch{CO2} ($>0.1$~bar).
    }
    \label{fig:photochem_fluxes}
\end{figure*}

\noindent
We thus use \textit{Atmos} with fixed molecular abundances in the lowest atmospheric layer as boundary conditions for \ch{H2}, \ch{CO2}, and \ch{CH4}.
We need to run the photochemical model under various values of these boundary conditions (and of luminosity) to simulate atmospheric evolution coupled with other processes.
This is computationally demanding, hence apart from \ch{H2}, \ch{CO2}, and \ch{CH4}, we set the boundary conditions for other gases to be the default fixed surface fluxes (Earth-like values) already set in \textit{Atmos} such that they are not dynamic variables in our coupled model.
For carbon monoxide (\ch{CO}), which is a dynamic variable of our coupled model through the modeling of its ocean chemistry (Methods Section \textit{Ocean-Atmosphere coupling}), we still use a constant boundary condition in \textit{Atmos} that is a surface deposition flux $v_{dep}=1.2\times10^{-4}$~cm~s$^{-1}$ \citep{kharecha2005coupled}.
This value for the deposition velocity of \ch{CO} is the default in \textit{Atmos} and corresponds to estimates based on the consumption of $\ch{CO}$ by organisms in the ocean, and is expected to be up to four orders of magnitude higher than abiotic estimates \citep{kharecha2005coupled,sauterey2020}.
Our model does not otherwise include any participation of biogenic fluxes to geochemical cycles.
Intuitively, this would lead to overestimate the rate of the CO-producing reactions, and underestimate that of CO-consuming reactions in the atmosphere, since imposing a higher sink of \ch{CO} at the surface forces equilibrium photochemical rates to produce more \ch{CO} than they would in a hypothetical case with only abiotic sinks.
In the atmosphere of the Archean Earth, \ch{CO} is produced by photolysis of \ch{CO2} \citep[][and references therein]{sauterey2020}, but \ch{CO} can also react with water to form \ch{CO2} \citep[among other molecules;][]{BarNun1983}.
This leads to non-trivial interactions between \ch{CO} and \ch{CO2} in our model, which do not appear in other models \citep{sauterey2020,kharecha2005coupled}.
Typical photochemical sources or sinks of carbon dioxide are found to be at most of the order of $10^4-10^6$~mol~yr$^{-1}$ (Figure~\ref{fig:photochem_fluxes}A), several orders of magnitude lower than typical values of volcanic outgassing or of weathering rates \citep[up to the order of $10^{13}$~mol~yr$^{-1}$;][and Figure~\ref{fig:schema}C]{krissansen2018constraining}.
Comparison between our model interpolation and the parameterization in the early-Earth model in \citet{sauterey2020} (which is based on the 1D version of the generic LMD global circulation model) reveals that the photochemical sink of \ch{CO2} in our model is lower by a factor 2.
At low \ch{CO2} pressures ($\approx 10^{-4}$~bar), our model predicts a \ch{CO2} source in atmospheric photochemistry (Figure~\ref{fig:photochem_fluxes}A), which is not predicted by the parameterization by \citet{sauterey2020}.
Thus, the particular value of the \ch{CO} deposition velocity is of negligible impact in our study.
However, future implementations of this model that may include \ch{CO} producing or consuming microbes should use atmospheric abundance boundary condition for this gas, as we do for \ch{CO2}, \ch{H2}, and \ch{CH4}.\\

\begin{figure}
    \centering
    \includegraphics[scale=1]{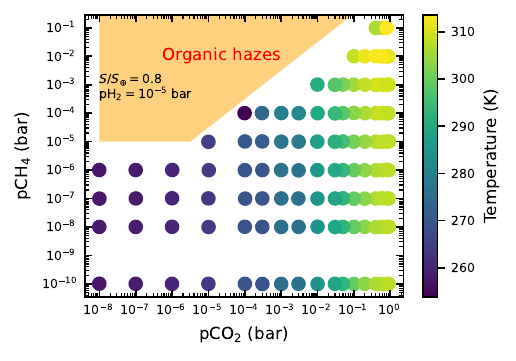}
    \caption{\textbf{Climate calculations in the $p\ch{CO2}$ versus $p\ch{CH4}$ space for $S/S_{\oplus}=0.8$ and $p\ch{H2}=10^{-5}$~bar.}
    The dots represent simulations made with the photochemistry/climate coupled model, and their color reflects the calculated surface temperature at radiative equilibrium.}
    \label{fig:climate_grid}
\end{figure}

\noindent
We ran simulations of the photochemistry-climate model over a slightly irregular grid of 8080 points sampled in the 4 dimensional volume defined by relative luminosity $S/S_{\oplus}$, and atmospheric abundance boundary conditions for $\ch{H2}$, $\ch{CO2}$ and $\ch{CH4}$.
The slight grid irregularity avoids known issues of high-dimensional linear interpolation using triangulation, namely discontinuities at evaluation points lying on shared triangulation edges caused by co-circular grid points \citep{barber1996quickhull}.
The boundaries of this grid were carefully chosen in order to keep computation time manageable while taking into account that this grid of simulations is intended to model the atmospheric evolution of habitable worlds with stable climates.
We sample boundary conditions ranging from $10^{-8}$ to $10^{-1}$ bars of $\ch{H2}$, $10^{-10}$ to $10^{-1}$ of $\ch{CH4}$ and from $10^{-8}$ to $0.95$ bars for $\ch{CO2}$ (Figure~\ref{fig:climate_grid}) and with a relative luminosity $S/S_{\oplus}$ varying from $0.4$ to $1.4$, corresponding to orbital distances from the Sun comprised between $0.85$ and $1.58$~au.
In addition, we set limits to our grid such that boundary conditions satisfy the conditions of $p\ch{H2}+p\ch{CH4}+p\ch{CO2}\leq 1$~bar, and $p\ch{CH4}/p\ch{CO2}<1$.
The latter condition prevents our simulated atmospheres from developing organic hazes \citep{zahnle1986photochemistry,kharecha2005coupled} which cause significant slow-down of climate calculations.
As shown later, some simulations of the dynamical evolution of the atmospheric composition may reach the border of the 4-D volume in which the atmosphere is modeled.
In such cases, the simulations are halted and the corresponding atmospheric scenario (likely hazes, or \ch{CO2} accumulation) is identified.\\

\begin{figure}
    \centering
    \includegraphics[width=8.7cm]{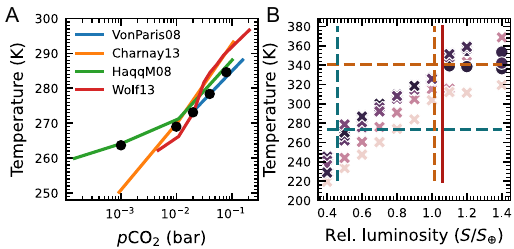}
    \caption{\textbf{A Comparison of habitable zone results}. The simulations are represented as crosses when they have converged and circles when they have not. In the cases that did not converge, the temperature retained and plotted is representative (the last recorded iteration of the climate simulation), but does not correspond to an atmosphere in radiative equilibrium.
    The color code of the simulations is shown in the legend and corresponds to the partial pressure of $\ch{CO2}$ forced constant at the bottom of the atmosphere.
    The blue dashed lines correspond to the outer edge of the HZ. The horizontal one is set at $T=0$°C, and the vertical one is set at the OHZ $S/S_{\oplus}=0.46$ value calculated in \citet{kopparapu2013habitable} for p\ch{CO2}=1~bar.
    The orange dashed lines represent the IHZ: horizontally at the surface temperature $T_S=340$~K at which the moist greenhouse or water loss limit is reached (Methods Section \textit{Atmospheric photochemistry and climate}), and vertically at the consensus luminosity at which this limit is reached $S/S_{\oplus}=1.015$ \citep{kopparapu2013habitable}.
    The solid red vertical line corresponds to the greenhouse runaway limit $S/S_{\oplus}=1.06$ \citep{kopparapu2013habitable}.
    \textbf{B Comparison of climate simulations} from the \textit{Atmos} model used here (black circles) and previously published simulations for an early Earth at $S/S_{\oplus}=0.8$ \citep[approximately 2.8 billion years ago,][]{charnay2013exploring,wolf2013hospitable,haqq2008revised,von2008warming}.
    These calculations assume fixed abundances p\ch{H2}=100ppm, and p\ch{CH4}=0.1ppm, which are the average values in our simulations with age $\tau \in [2,3]$~Gyrs and $S/\ssun \in [0.75,0.85]$.}
    \label{fig:clima_comparisons}
\end{figure}

\noindent
It is important to note here that we assumed the same type of vertical profile of water vapor across the entire range of our atmospheric photochemistry and climate simulations.
The vertical profile of water that we used to initialize the climate-photochemistry simulations is described by \citet{manabe1967thermal}, and assumes a non water-saturated troposphere.
This water vapor vertical profile matches empirical expectations for the atmosphere of the modern Earth \citep{catling2017atmospheric}.
It is unknown whether such a profile is more or less realistic than assuming a water-saturated troposphere as commonly done to infer the water loss limit \citep{kopparapu2013habitable}.
In consequence, our simulations may underestimate the surface temperature of planets that are closer to the inner edge of the HZ, thus deviating from results of \citet{kopparapu2013habitable}, as shown in Figure~\ref{fig:clima_comparisons}A.
Furthermore, our assumption of a 1 bar atmosphere prevents modeling $\ch{CO2}$-rich dense atmospheres ($>1$~bar) that could remain temperate at relative luminosity values as low as $S/S_{\oplus}=0.34$ \citep{kopparapu2013habitable}.
Nonetheless, these cases (where \ch{CO2} accumulates above 1 bar) are encountered in $\sim7$\% of all simulations, and may still be used in our statistical analyses.
Further exploration of changes in water saturation is beyond the scope of this first study, but future work should allocate effort into relaxing that assumption.\\

\noindent
Within the radiative equilibrium regime, we observe that the temperatures calculated in our climate simulations are in relatively good agreement with previously published results (Figure~\ref{fig:clima_comparisons}B).
1D models appear to overestimate the warming effect of low partial pressures of $\ch{CO2}$, and underestimate it at higher abundances.
These differences are well documented \citep{Fauchez2021}.
Three-dimensional Global Climate Models (GCMs) are expected to constitute a reference, and the differences mentioned above are to be kept in mind when interpreting the results of our simulations relying on the \textit{Atmos} climate model \citep[see][for a comprehensive comparison between different atmosphere models]{Fauchez2021}.

\noindent
In order to use the climate and photochemical calculations to integrate the set of differential equations summarized by Equation (\ref{eq:der_gen}), the result of atmospheric simulations are interpolated \citep[using the linear ND interpolator from SciPy;][]{virtanen2020scipy} in the 4-D space comprising the grid (utilizing the $\log_{10}$ of partial pressures) in order to obtain the following continuous functions:
{\footnotesize
\begin{equation}
    \begin{array}{rcl}
        (T_S,A_P) & = & F_1(p\ch{H2},p\ch{CO2},p\ch{CH4},S/S_{\oplus}) \\
        (\Phi_{\ch{H2}},\Phi_{\ch{CO2}},\Phi_{\ch{CH4}},\Phi_{\ch{CO}}) & = & F_2(p\ch{H2},p\ch{CO2},p\ch{CH4},S/S_{\oplus})
    \end{array}
    \label{eq:atmosphere_approx}
\end{equation}}
where $T_S$ (K) is the surface temperature, and $A_P$ the planetary albedo.
$\Phi_i$ (molecules~cm$^{-2}$~s$^{-1}$) are photochemical fluxes.

\subsection{Ocean-Atmosphere coupling}
Next, atmospheric composition is influenced by the exchange of gases across the ocean's surface.
Previous models implicitly assume that the ocean and the atmosphere are at solubility equilibrium, i.e. that the concentration of molecule $\ch{X}_i$ in the ocean follows Henry's law $[\ch{X}_i]=\alpha_{\ch{X_i}} p\ch{X_i}$, with $\alpha_{\ch{X_i}}$ ($\text{mol}~\text{L}^{-1}~\text{bar}^{-1}$) the Henry coefficient for $\ch{X_i}$, and $p\ch{X_i}$ the gas partial pressure of $\ch{X_i}$ in the atmosphere \citep[bar; \textit{e.g.}][]{krissansen2017constraining}.
Instead of assuming thermodynamic equilibrium between the ocean and the atmosphere, we explicitly describe the dynamics of the concentrations of relevant molecules in the ocean governed by subaquatic volcanic outgassing, and carbonate precipitation and dissolution.
These dynamics include a net flux to or from the atmosphere due to the ocean and atmosphere not being at thermodynamic equilibrium.
We then calculate the steady-state of the ocean's composition by assuming that the atmospheric composition remains constant on the timescale over which this equilibrium is reached.
As a result, we obtain the flux across the ocean/atmosphere interface as a function of atmospheric composition.

\subsubsection{Molecular flux across the ocean/atmosphere interface}
The exchange rate between the ocean and the atmosphere is described by Equation (\ref{eq:Foc}), corresponding to the stagnant boundary layer (SBL) model \citep{kharecha2005coupled}:
\begin{equation}
    \label{eq:Foc}
    \mathrm{F}_{oc}(\ch{X_i}) = v_{\ch{X_i}}(\ch{[X_i]} - \alpha_{\ch{X_i}} p\ch{X_i})\times C
\end{equation}
where $\mathrm{F}_{oc}$ ($\text{molecules}~\text{cm}^{-2}~\text{s}^{-1}$) is the flux of molecule $\ch{X}_i$ across the atmosphere-ocean interface (counted positive for flux to the atmosphere) for the chemical species $\ch{X}_i$, $v_{\ch{X}_i}$ is a piston velocity across the stagnant boundary layer ($\text{cm}~\text{s}^{-1}$, for a $40~\mu\text{m}$ thick SBL), $\alpha_{\ch{X}_i}$ is the Henry coefficient ($\text{mol}~\text{L}^{-1}~\text{bar}^{-1}$), $p\ch{X}_i$ is the gas partial pressure in the atmosphere (bar), $[\ch{X}_i]$ is the concentration of $\ch{X}_i$ in the ocean ($\text{mol}~\text{L}^{-1}$) and $C=\mathrm{A}/10^{3}~\text{molecules}~\text{mol}^{-1}/(\text{cm}^3~\text{L}^{-1})$ with $\mathrm{A}=6.02\times 10^{23}$~mol$^{-1}$ the Avogadro number.\\

\noindent
To track concentration change in the ocean, it is more useful to rearrange some elements of Equation (\ref{eq:Foc}) into a diffusion constant
\begin{equation}
    \label{eq:diffusion}
    D_{\ch{X_i}} = \frac{v_{\ch{X_i}}}{z_{oc}} ~ (\text{s}^{-1})
\end{equation}
where $z_{oc}$ is the ocean depth (in cm).
The volume of the ocean is set constant using $M_{o} = 3\times 10^{21}$~kg and $\rho_{oc}=1000$~kg~m$^{-3}$ in all our simulations.
However, the ocean surface is not set constant, as we allow variable emerged land surface.
Thus, the depth of the ocean $z_{oc}$ is calculated using
\begin{equation}
    z_{oc} = \frac{V_{oc}}{S_p (1-f_{land})}
\end{equation}
where $S_p=5.1\times 10^{18}$~cm$^{-2}$ is the total surface of the planet, and $f_{land}$ is the fraction of the surface of emerged land (not necessarily continental crust) so that the ocean surface $S_{oc}=(1-f_{land})S$.
Last, $\mathrm{F}_{oc}$ ($\text{molecules}~\text{cm}^{-2}~\text{s}^{-1}$) is converted into $\mathrm{F}_{oc}^{\prime}$ (mol~yr$^{-1}$):
\begin{equation}
    \label{eq:ocatm_exchange}
    \mathrm{F}_{oc}^{\prime} = D_{\ch{X}_i}(\alpha_{\ch{X}_i}p\ch{X_i} - [\ch{X}_i])\times (3600\times24\times365.25)~.
\end{equation}

\subsubsection{Carbon monoxide ocean chemistry}
Following the Appendix 3 of \citet{kharecha2005coupled}, we parameterize the kinetics of the photochemical conversion, in the upper ocean, of \ch{CO} to formate (\ch{HCOO-}) and subsequently to acetate following
\begin{equation}
\label{eq:cokinetics}
    \ch{CO} + \ch{OH-} \xrightleftharpoons[k_2]{k_{hyd}} \ch{HCOO-} (+ \cdots) \xrightarrow{k_3} \ch{CH3COO-} (+ \cdots) \space.
\end{equation}
The steady-state of these reactions (equation~\ref{eq:cokinetics}) coupled with the rate of exchange with the atmosphere (equation~\ref{eq:ocatm_exchange}) results leads to
\begin{equation}
    \label{eq:steady_co}
    \begin{array}{r c l}
    [\ch{CO}]    & = &\frac{\alpha_{\ch{CO}}p\ch{CO} D_{\ch{CO}}(k_2+k_3)}{k_3 k_{hyd} [\ch{OH-}]+(k_2+k_3)D_{\ch{CO}}} \\
    
    [\ch{HCOO-}] & = &\frac{\alpha_{\ch{CO}} p\ch{CO} D_{\ch{CO}} k_{hyd} [\ch{OH-}]}{k_3 k_{hyd} [\ch{OH-}]+(k2+k3) D_{\ch{CO}}} \space
    \end{array}
\end{equation}
where $[\ch{OH-}]$ is derived from the pH (variable in our model) and the ionic product of water in standard conditions $\ch{pK_e}=\ch{pH}+\ch{pOH}=14$.

\subsubsection{Carbon cycle in a coupled ocean-atmosphere model}
Continental weathering transfers atmospheric (and crustal) carbon to the ocean.
This occurs as the products of the weathering of continental crust minerals are transported in rivers:
{\footnotesize
\begin{equation}
\label{reaction:contweath}
\begin{array}{rcl}
    \ch{CO_2} + \ch{H_2O} + \ch{CaCO_3} & \longrightarrow & \ch{Ca^{2+}} + 2 \ch{HCO_3^-} \\
    2\ch{CO_2} + \ch{H_2O} + \ch{CaSiO_3} & \longrightarrow & \ch{Ca^{2+}} + 2 \ch{HCO_3^-} + \ch{SiO_2} \space .
\end{array}    
\end{equation}}
Then, carbonates can precipitate in the ocean (at a pH-dependent rate noted $P_o$ in mol~yr$^{-1}$; equation~\ref{reaction:prec}), of which a certain fraction returns to the crustal reservoir through diagenesis
\begin{equation}
\label{reaction:prec}
    \ch{Ca^{2+}} + 2 \ch{HCO_3^-} \longrightarrow \ch{CO_2} + \ch{H_2O} + \ch{CaCO_3} \space .
\end{equation}
The rate of the reactions in Equation (\ref{reaction:contweath}) are assumed to have increased with the global mean surface temperature of the Earth, hence providing a negative feedback loop on climate through the abundance of atmospheric carbon \citep[Figure~\ref{fig:schema}B;][]{walker1981negative,krissansen2017constraining}.
It has been assumed that other terrestrial planets could undergo similar climate stabilization.
This is used to explain the inner edge of the HZ being set by the greenhouse effect of water vapor, and that the maximum partial pressure of \ch{CO2} (before condensation) setting the traditional outer edge of the HZ \citep{kasting1993habitable,kopparapu2013habitable}.\\

\noindent
A second temperature-sensitive process that can transfer atmospheric carbon to the crustal reservoir is seafloor weathering \citep{krissansen2017constraining,gillis2011secular,alt1999uptake}.
This process does not rely on the alteration of continental crust, but rather on the dissolution of the basaltic seafloor occurring at off-axis hydrothermal circulation, releasing cations such as \ch{Ca^2+} hence promoting reaction \ref{reaction:prec} (by increasing $[\ch{Ca^2+}]$ and carbonate alkalinity through ionic balance; see Figure~\ref{fig:schema}B).

\noindent
We largely reproduce the model of carbonate chemistry and the carbonate-silicate weathering cycle in \citet{krissansen2017constraining} with only slight modifications to allow the bulk ocean and the atmosphere to be in disequilibrium \citep[unlike][where the ocean and atmosphere are treated as at equilibrium]{krissansen2017constraining,lehmer2020carbonate}.

\noindent
Following \citet{krissansen2017constraining} we track the concentration of dissolved carbon (DIC) and carbonate alkalinity (ALK; both defined in Equation \ref{eq:DICALK}) in the upper and deep ocean that mix at constant rate $J=6.8\times 10^{16}$~kg~yr$^{-1}$.
\begin{equation}
    \label{eq:DICALK}
    \begin{array}{rcl}
        \ch{DIC} & = & \ch{[CO_2]_{aq}} + \ch{[HCO_3^-]} + \ch{[CO_3^{2-}]} \\
        \ch{ALK} & = & \ch{[HCO_3^-]} + 2\ch{[CO_3^{2-}]}~.
    \end{array}
\end{equation}
The overall dynamics of the carbonate system in the ocean (subscript $o$) and deep ocean (subscript $d$) follows:
{\footnotesize
\begin{equation}
\label{main:eq:carboncycle}
    \begin{array}{r c l}
        \dfrac{\mathrm{d} C_o}{\mathrm{d}t} & = & \overbrace{D_{\ch{CO2}}(\alpha_{\ch{CO2}}p\ch{CO2}-[\ch{CO2}])}^{\text{atmosphere exchange}} - \overbrace{\frac{J}{M_o} (C_o - C_d)}^{\text{ocean mixing}}\\
         & & + \frac{S_p}{M_o}(\overbrace{f_{land} F_{cont}}^{\text{weathering}} + \overbrace{(1-f_{land})\times (F_{out}-P_o)}^{\text{oc. outgassing and precipitation}})\\[10pt]
         \dfrac{\mathrm{d} \ch{ALK}_o}{\mathrm{d}t} & = & \frac{S_p f_{land}}{M_o} (F_{cont} - 2 P_o) - \frac{J}{M_o}(\ch{ALK_o} - \ch{ALK_d}) \\[10pt]
         \dfrac{\mathrm{d} C_d}{\mathrm{d}t} & = & \frac{J}{M_d} (C_o - C_d) + (1-f_{land})\frac{S_p}{M_d} (F_{out} - P_d)\\[10pt]
         \dfrac{\mathrm{d} \ch{ALK}_d}{\mathrm{d}t} & = & \frac{J}{M_d}(\ch{ALK_o} - \ch{ALK_d})\\
          & & + 2 (1-f_{land})\frac{S_p}{M_d} (F_{diss} - P_d)\\[10pt]
          \dfrac{\mathrm{d} C_d}{\mathrm{d}t} & = & \frac{J}{M_d}(C_o - C_d) - (1-f_{land})\frac{S_p}{M_d} P_d
    \end{array}
\end{equation}}
where $C_o$ and $C_d$ are the dissolved inorganic carbon (DIC) concentrations in the upper and deep ocean respectively and likewise $\ch{ALK}_o$ and $\ch{ALK}_d$  describe the carbonate alkalinity.
$D_{\ch{CO2}} \approx 0.25$~yr$^{-1}$ is the diffusion constant of $\ch{CO2}$ in the SBL model (computed using a piston velocity $v_{\ch{CO2}}=4.8\times 10^{-3}$~cm~s$^{-1}$, see Table~\ref{table:parameters}).
$M_o=3\times 10^{21}$~kg is the upper ocean mass (Table~\ref{table:parameters}) and $M_d=0.01\times M_o$.
The continental weathering flux
\begin{equation}
    \label{eq:fcont}
    F_{cont} = 2F_{carb}+2F_{sil}
\end{equation}
is the sum of the carbonate and silicate weathering fluxes (mol~yr$^{-1}$~cm$^{-2}$).
$S_p=5.1 \times 10^{18}$ cm$^2$ is the planetary surface area (fixed to Earth's value).\\

\noindent
Weathering rates, as well as those of carbonate precipitation in the upper and deep ocean ($P_o$ and $P_d$ in mol~yr$^{-1}$~cm$^{-2}$) have been given parameterized equations in \citet{krissansen2017constraining}:
\begin{equation}
    \label{eq:fcarb}
    F_{carb}\times S_p/f_{land} = \frac{1}{f_{land}^{ME}} F_{carb}^{ME} \left( \frac{p\ch{CO2}}{p\ch{CO2}^{ME}} \right)^{\chi} e^{\frac{T_s-T_{ME}}{T_e}}
\end{equation}
where ${f}_{land}^{ME}=0.29$ denotes the surface area fraction of emerged land of the modern Earth, $F_{carb}^{ME}$ is the value of the modern Earth carbonate weathering flux, $p\ch{CO2}^{ME}$ is the modern Earth value of p\ch{CO2}, $T_{ME}$ the modern Earth's surface temperature, $\chi$ is the coefficient setting the dependency of weathering rate on p\ch{CO2}, and $T_e$ is the e-folding temperature of the weathering reaction.
Similarly,
\begin{equation}
    \label{eq:fsil}
    F_{sil}\times S_p/f_{land} = \frac{1}{f_{land}^{ME}} F_{sil}^{ME} \left( \frac{p\ch{CO2}}{p\ch{CO2}^{ME}} \right)^{\alpha_s} e^{\frac{T_s-T_{ME}}{T_e}}
\end{equation}
The value of these parameters is given in Appendix Tables~\ref{table:initialization},\ref{table:parameters}.\\

\noindent
Carbonate precipitation occurring above the continental shelf results in effective carbon removal from the ocean-atmosphere surface.
In contrast, pelagic carbonate precipitation (occurring in the open ocean), can result in carbonate sinking to depths exceeding the so-called carbonate compensation depth (CCD).
At this depth, increased hydrostatic pressure allows carbonates to resolubilize.
This results in a different net surface specific rate of precipitation in above-shelf and pelagic ocean.
\citet{krissansen2017constraining} propose a functional form of the pelagic precipitation rate, in which the area of the ocean that is above the CCD is parameterized.
The expression of the effective precipitation rate is
\begin{equation}
    P_o\times S_p = k_{shelf}(\Omega_o-1)^{n_{carb}} + k_{pelagic}\Omega_o^{2.84}
\end{equation}
where $k_{shelf} = 1.03 \times 10^{13}$~mol~yr$^{-1}$ and $k_{pelagic}=1.40 \times 10^{12}$~mol~yr$^{-1}$ are reference values of above-shelf and pelagic precipitation \citep{krissansen2017constraining}.
The coefficients $n_{carb}$ (sampled from the range given in Appendix Table \ref{table:initialization}) and $n_2=2.84$ set the sensitivity of precipitation rates to the saturation state $\Omega$, which is detailed in Equation (\ref{eq:omega}).\\

\noindent
The deep ocean also has a precipitation term:
\begin{equation}
    P_d\times S_p = k_{p}(\Omega_d-1)^{n_{carb}}
\end{equation}
where $k_p=5.12 \times 10^{11}$~mol~yr$^{-1}$.\\

\noindent
The saturation state $\Omega$ is given in compartment $c$ (upper or deep ocean) by
\begin{equation}
\label{eq:omega}
    \Omega_c = \frac{[\ch{Ca^{2+}}]_c[\ch{CO3^{2-}}]_c}{K_{sol}}
\end{equation}
where $K_{sol}$ ($\text{mol}^{2}~\text{kg}^{-2}$) is the solubility product of carbonates (see Appendix \ref{chemeq_appendix}).
The concentration of calcium cations is obtained by \citep{krissansen2017constraining}:
\begin{equation}
    \label{eq:cadeuxplus}
    [\ch{Ca^2+}](t) = [\ch{Ca^2+}](0) + \dfrac{\ch{ALK}-\ch{ALK}_0}{2}~.
\end{equation}
The concentration $[\ch{CO3^{2-}}]$ is obtained from resolving the pH-dependent carbonate equilibrium.
We assume that carbonate equilibrium is instantaneous and we follow \citet{krissansen2017constraining} and \citet{krissansen2018constraining} in tracking carbonate alkalinity (ALK) and total dissolved inorganic carbon (DIC) as dynamical quantities in order to solve for the pH and consequently the carbonate equilibrium :
\begin{equation}
\label{eq:carb_system}
\begin{array}{rcl}
    \ch{DIC} & = & \ch{[CO_2]_{aq}} + \ch{[HCO_3^-]} + \ch{[CO_3^{2-}]} \\
    \ch{ALK} & = & \ch{[HCO_3^-]} + 2\ch{[CO_3^{2-}]} \\
    \ch{[HCO_3^-]} & = & \frac{\ch{[CO_2]_{aq} K_1}}{[H^+]} \\ 
    \ch{[CO_3^{2-}]} & =&  \frac{\ch{[HCO_3^-] K_2}}{[H^+]} \\ 
    \end{array}
\end{equation}
where $K_1$ is the equilibrium constant for $\ch{(CO2)_{aq}} \rightleftharpoons \ch{HCO3-}$ and $K_2$ is the equilibrium constant for $\ch{HCO3-} \rightleftharpoons \ch{CO_3^{2-}}$.
These equilibrium constants are given as functions of temperature in Appendix \ref{chemeq_appendix}.\\

\noindent
By combining equations in Equation (\ref{eq:carb_system}), we obtain :
\begin{equation}
\label{main:eq:dic_alk}
    \left\{
    \begin{array}{rcl}
        \ch{DIC} & = & \ch{[CO_3^{2-}]} \left( \frac{\ch{[H^+]^2}}{K_1 K_2} + \frac{\ch{[H^+]}}{K_2} + 1 \right) \\
        \ch{ALK} & = & \ch{[CO_3^{2-}]} \left( \frac{\ch{[H^+]}}{K_2} + 2 \right)
    \end{array}
    \right.
\end{equation}
ultimately yielding a quadratic Equation that can be solved for $\ch{[H^+]}$:
\begin{equation}
\label{eq:pH}
    \ch{[H^+]^2} \frac{\ch{ALK}}{K_1 K_2} + \ch{[H^+]} \frac{\ch{ALK}-\ch{DIC}}{K_2} + \ch{ALK} - 2 \ch{DIC} = 0
\end{equation}
From this, $[\ch{CO2}]_{aq}$, $[\ch{HCO3-}]$, $[\ch{CO_3^{2-}}]$, and the precipitation rate of carbonate follow.\\

\noindent
Temperature-dependent basalt dissolution in the seafloor (occurring at rate $F_{diss}$, mol~yr$^{-1}$~cm$^{-2}$) releases calcium ions that increase carbonate alkalinity as a result of ionic balance and ultimately promotes carbonate precipitation \citep[see Figure~\ref{fig:schema}B;][]{alt1999uptake,gillis2011secular,krissansen2017constraining}.
We use the expression from \citet{krissansen2017constraining}:
\begin{equation}
    \label{eq:fdiss}
    \begin{array}{rcl}
    F_{diss} \times S_p & = & k_{diss} \left(\frac{F_{out}}{F_{out}^{ME}}\right)^{\beta}\\
     & & \times \exp{\left(-\frac{E_{bas}}{R T_p}\right)} \left(\frac{[\ch{H+}]_p}{[\ch{H+}]_p^{ME}}\right)^{\gamma}
    \end{array}
\end{equation}
where $k_{diss}$ (mol~yr$^{-1}$) is set so that the expression given here equals $F_{diss}^{ME}$, $F_{out}$ is the outgassing rate of $\ch{CO2}$ (which is a parameter of the simulations, see below) and $F_{out}^{ME}$ (mol~yr$^{-1}$) is the present value of volcanic outgassing of $\ch{CO2}$.
The rate of volcanic outgassing is assumed to scale with the production of fresh basalt which limits the supply of weatherable material.
$E_{bas}$ (kJ~mol$^{-1}$) is the activation energy of basalt dissolution, $T_p=a_{grad} T+(274.037-a_{grad} \times 285)+9$ the temperature of pore-space water (K), $\beta$ and $\gamma$ set the sensitivity of the basalt dissolution rate to basalt supply and pH respectively.
The value of these parameters is sampled at random from credible ranges established by \citet{krissansen2017constraining} and recalled in Appendix Table~\ref{table:initialization}.
$[\ch{H}^+]_p^{ME}$ is the modern value of pH in Earth's ocean, $-\log_{10}[\ch{H}^+]_p^{ME} = 8.2$.
These parameters are recalled in Appendix Table~\ref{table:parameters}.

\subsubsection{Ocean steady-state}
In the upper ocean, concentration $[\ch{X}_i]$ generally follows
\begin{equation}
\label{eq:diff_ochem}
\begin{array}{rcl}
     \dfrac{\mathrm{d}[\ch{X}_i]}{\mathrm{d}t} & = & D_i (\alpha_i \ch{pX_i} - [\ch{X}_i]) - \frac{J}{M_o}([\ch{X}_i] - [\ch{X}_i]_d) \\
     \dfrac{\mathrm{d}[\ch{X}_i]_d}{\mathrm{d}t} & = & \frac{1}{M_d} \left[ J([\ch{X}_i] - [\ch{X}_i]_d) + F_{i,geol} \right]
\end{array}
\end{equation}
where $D_i$ is the diffusion mixing constant obtained from the stagnant boundary layer model (Equation \ref{eq:diffusion}), $\alpha_{\ch{X}_i}$ is the Henry's law coefficient for $\ch{X}_i$, $p\ch{X}_i$ its partial pressure in the lower atmosphere, $J$ the ocean mixing constant, $[\ch{X}_i]_d$ the concentration of $i$ in the deep ocean and $F_{i,geol}$ in mol~yr$^{-1}$ summarizes the geological (outgassing, serpentinization) fluxes of $\ch{X}_i$ assumed to occur in the deep ocean.\\

\noindent
Assuming a separation of timescales between ocean chemistry and change of the atmospheric composition, Equation (\ref{eq:diff_ochem}) can be solved for steady-state $[\ch{X}_i]^*$ (upper ocean) assuming constant partial pressure p\ch{X_i}:
\begin{equation}
    \label{eq:ochem_ss}
    [\ch{X}_i]^* = \alpha_{\ch{X}_i} \ch{pX}_i + \dfrac{(1-f_{land})F_{i,geol}}{D_i M_o}
\end{equation}
Using Equation (\ref{eq:ocatm_exchange}), the general case steady-state value of the flux to or from the atmosphere due to interaction with the ocean simplifies into $\mathrm{F}_{oc}([\ch{X}_i]^*) = (1-f_{land})F_{i,geol}$.
In the specific case of \ch{CO2}, however, this flux must be explicitly calculated using Equation (\ref{eq:Foc}) with the value for $[\ch{CO2}]_{aq}$ obtained by solving Equation (\ref{eq:pH}) and using the resulting value for \ch{pH} in Equation (\ref{eq:carb_system}).
Similarly, the flux of \ch{CO} is obtained by plugging Equation (\ref{eq:steady_co}) into Equation (\ref{eq:Foc}).

\subsection{Subaerial volcanic outgassing}
Volcanic outgassing of \ch{CO2} ($F_{out}$, mol~yr$^{-1}$) is the principal source of atmospheric carbon in our modeled atmospheres.
The intensity of volcanic outgassing also sets the intensity of seafloor weathering through supply of weatherable minerals (Equation \ref{eq:fdiss}).
Here, although each individual simulation assumes a constant rate of volcanic outgassing (as part of the parameter vector $\boldsymbol{\theta}$), we sample the value of this parameter \textit{via} a planetary age variable that is denoted by $\tau$ (Gyr).
The value of $\tau$ is sampled uniformly in the range 0.5 to 4.5~Gyr (Appendix Table~\ref{table:initialization}).\\

\noindent
If the simulation is in the ML regime, we calculate the value of the \ch{CO2} outgassing rate from planetary age following the parameterization used in \citet{krissansen2018constraining}:
\begin{equation}
    \label{eq:agefout_ML}
    F_{out}^{ML} = F_{out}^{ME}\times Q^{m}
\end{equation}
with
\begin{equation}
    \label{eq:foutQ}
    Q    = \left(1-\frac{(4.5-\tau)}{4.5}\right)^{-n_{out}}~.
\end{equation}
The values of power law parameters $m$ and $n_{out}$ are sampled from ranges following \citet{krissansen2017constraining} and recalled in Appendix Table~\ref{table:initialization}.
$F_{out}^{ME}$ (mol~yr$^{-1}$) is the present-day value of \ch{CO2} volcanic outgassing (sampling range is given in Appendix Table~\ref{table:parameters}).\\

\noindent
If the simulation is in the SL regime, we use the relation between planetary age and volcanic outgassing from \citet{dorn2018outgassing}.
The work by \citet{dorn2018outgassing} is based on inference of interior structure and composition of super-Earths and subsequent simulation of convection and melting of the inferred mantle; outgassing is assumed to be directly proportional to the calculated melting.
We fit a sigmoid function $f(\tau) = K/(1+a e^{-r x})$ to the data from \citet{dorn2018outgassing} for 1 Earth-mass planets in order to estimate the values of parameters $K,a,r$ using the least mean square method.
This gave parameter values $a=481.28$, $r=5.12$, and $K=42.57$.
Lastly, the outgassing flux is expressed as the derivative of the accumulated atmospheric \ch{CO2} with appropriate conversion to units of mol~yr$^{-1}$:
\begin{equation}
    \label{eq:agefout_SL}
    F_{out}^{SL}(\tau) = 10,000 S_P \frac{101,300}{\frac{M_{\ch{CO}}}{1,000} g}\frac{a K r e^{-r \tau}}{(1+ a e^{-r \tau})^2}
\end{equation}
where $g=9.81$~N~kg$^{-1}$ the gravitational acceleration for a 1 Earth mass planet, $M_{\ch{CO2}}=44$~g~mol$^{-1}$ the molar mass of carbon dioxide, and $S_P$ is the planetary surface area of a 1 Earth radius planet (Appendix Table~\ref{table:parameters}).\\

\noindent
The values for the outgassing of methane ($Serp$, mol~yr$^{-1}$) and dihydrogen ($Volc$, mol~yr$^{-1}$) are expressed relative to the \ch{CO2} volcanic outgassing:
\begin{equation}
    \label{eq:serp}
    Serp = Serp^{ME}\frac{F_{out}}{F_{out}^{ME}}
\end{equation}
for methane, and
\begin{equation}
    \label{eq:volc}
    Volc = Volc^{ME}\frac{F_{out}}{F_{out}^{ME}}
\end{equation}
for dihydrogen.
The modern values $Serp^{ME}=10^{11}$~mol~yr$^{-1}$ and $Volc^{ME}=5.35 \times 10^{12}$~mol~yr$^{-1}$ \citep{sauterey2020} are recalled in Appendix Table~\ref{table:parameters}.\\

\noindent
Our model makes the approximation that serpentinization and volcanic outgassing occur uniformly on emerged land and under the ocean, such that the rate of change of $p\ch{X}_i$ due to outgassing scales with $F_{i,geol}\times f_{land}$.

\subsection{Forward-in-time modeling of atmosphere composition dynamics}
Methods Sections \textit{Atmospheric photochemistry and climate} and \textit{Subaerial volcanic outgassing} expand the components of Equation (\ref{eq:der_gen}).
Its form can now be detailed by adding the atmospheric photochemistry component $\boldsymbol{\Phi}$ (Equation \ref{eq:atmosphere_approx}), the ocean component (Equation \ref{eq:Foc}) and a subaerial volcanic component $f_{land} F_{i,geol}$
\begin{equation}
    \label{main:eq:ocderivative}
    \begin{array}{rcll}
        \dfrac{\mathrm{d} p \ch{X_i}}{\mathrm{d}t} & = & \frac{P_{tot}}{n_{tot}} S_p \left[ \right. & \frac{1}{\mathrm{A}} \left\{ (1-f_{land}) F_{i,oc}+\Phi_{\ch{X}_i} \right\} \\
         & & & \left. + f_{land} F_{i,geol}  \right]
    \end{array}
\end{equation}
where $n_{tot}$ and $P_{tot}$ are the total number of air moles in the atmosphere and the total atmospheric pressure, given in Appendix Table~\ref{table:parameters}.
To the variables describing partial pressures in the atmosphere, we add values of carbonate alkalinity and dissolved carbon concentration in the surface and deep ocean, for which the timescale separation assumption is not possible ($\ch{ALK}_o$, $\ch{ALK}_d$, $C_o$ and $C_d$, see Methods Section \textit{Ocean-atmosphere coupling}) to compose the full system to be integrated (Equation \ref{eq:der_gen}).
Integration of this system of differential equations is performed by using the Backward Differentiation Formula (BDF) of variable order.
This implicit integration scheme is particularly suited for stiff problems, where processes operate with very different timescales such as ours.
However, the BDF integrator requires that the roots of an equation are numerically solved at each timescale, which may require evaluation of $F_{\mathbf{\theta}}(\mathbf{y})$ (in equation \ref{eq:der_gen}) outside of its domain of definition limited by the atmospheric grid.
In addition, the BDF requires calculation (in our case numerical) of the Jabobian matrix of the differential system.
The linear interpolation applied to the atmosphere grid makes molecular photochemical fluxes non differentiable at grid points, possibly corresponding to system states at which the Jacobian matrix is not defined.
For those reasons, numerical integration using the BDF may sometimes fail, even when the simulation trajectory remains within our grid of atmospheric calculations.
When such case arises, we switch to an explicit integrator (Runge-Kutta of order 4, with order 5 error calculation; RK45) for 10,000 simulation years after which the integrator is switched back to BDF.
It is to be noted that the RK45 integrator requires at each time step evaluation of $\dot{y}=f(t,y)$ at system states away from the solution $y(t)$, such that it is still possible for the RK45 integrator to numerically fail even when the solution would remain in the atmospheric grid validity domain.
However, we find switching to RK45 to help integration progress past system states where BDF fails in a vast majority of cases, and numerical failure was ultimately rare, as discussed later.

Simulations are run until the following criterium for convergence to steady-state is met for all elements of $\dot{\mathbf{y}}$:
\begin{equation}
    \dot{\mathbf{y}} < \mathbf{a} + r \times \mathbf{y}
\end{equation}
where $r=10^{-9}$ (such that the relative tolerance term corresponds to a rate of change of 1 per mil per million years), and the elements of $a$ corresponding to ocean concentration variables are set to $10^{-3}$~$\mu$mol~Myr$^{-1}$, and those corresponding to atmospheric concentrations are set to $0.1$ ppb~Myr$^{-1}$.
Initial conditions used for integration are described in Methods Section \textit{Initialization}.

\subsection{Initialization}
Different initial conditions and parameter values define the variability in our sample.
The initial state and parameter values of each simulation are sampled in distributions given in Appendix Table~\ref{table:initialization}.\\

\noindent
Rather than initializing the simulation with values of $\ch{ALK}$ and $\ch{DIC}$, that are time-dependent variable in the ODE described in Methods Section \textit{Forward-in-time modeling of atmosphere evolution}, we use initial $p\ch{H}$ and $p\ch{CO2}$ to calculate the initial values of $\ch{ALK}$ and $\ch{DIC}$ in the ocean using Equation (\ref{main:eq:dic_alk}), assuming that $[\ch{CO2}]$ is at equilibrium with the atmosphere:
\begin{equation}
    [\ch{CO2}]_0 = \alpha_{\ch{CO2}} \times p\ch{CO2}|_0
\end{equation}
The $\ch{ALK}$ and $\ch{DIC}$ in the deep ocean are then calculated at steady state with the upper ocean, only taking into account ocean mixing, which results in equal values in the deep and upper ocean (thus canceling out the mixing rate term $J$).
Together, the initial atmospheric composition sampled from Appendix Table~\ref{table:initialization} and the calculated initial alkalinity and dissolved carbon form the initial values of the time dependent variables in the simulation, and numerical integration can be started from there.

\subsection{Simulations batch}
We launched the simulations of climate and atmospheric composition for 40,000 randomly sampled initial conditions and parameters $\tau$ (age) and $f_{land}$, as well as other parameters of the carbon cycle model, especially those shaping the response of geological carbon fluxes to global temperature and atmospheric carbon dioxide, (see Appendix Table~\ref{table:initialization}) for each interior convection regime.

For ML simulations, continental weathering ($F_{cont}$) follows Equation (\ref{eq:fcont}), for SL simulations, continental weathering is set to $F_{cont}=0$.
Some simulations did not reach steady-state conditions as defined by our criterion (equation \ref{eq:steady_co}).
Among them, some were stopped for well-identified reasons.
Simulations that reach the maximum \ch{CO2} atmospheric partial pressure allowed by the climate-photochemistry grid (i.e. for which radiative equilibrium cannot be found consistently with imposed boundary conditions; this limit is typically between 0.01 and 0.95 bar depending on the luminosity), and that have a positive rate of change of $p\ch{CO2}$ are halted and labeled as `\ch{CO2} accumulation' cases.
This case was encountered in 203 ML simulations (0.5\%) and 5445 SL simulations (13.6\%).
Other simulations however, may fail to reach steady-state for a variety of reasons that include oscillations or wandering outside of the validity domain of our interpolated photochemical and climate model, or to edge cases where parameterized modeling component are also no longer valid (e.g., extreme ocean acidification).
These only occurred rarely in our sample (2-4\% of all simulations depending on the interior convection regime scenario), and out of caution we discarded them from our analyses.
In total, we use the dataset composed of simulations that have either reached steady-state, or that have been halted for \ch{CO2} accumulation or wipeout, corresponding to 39318 ML (98.3\%) and 38590 SL (96.5\%) simulations.

\section{Results}
\subsection{Effect of luminosity on atmospheric \ch{CO2}.}
For terrestrial planets in the HZ of their host star, continental weathering is expected to result in a negative p\ch{CO2}-luminosity correlation \citep{lehmer2020carbonate}.
This correlation is due to the carbonate-silicate weathering cycle, which is sensitive to temperature in the model we use.
At a given atmospheric \ch{CO2}, higher luminosity would cause surface temperature to rise (see Figure~\ref{fig:climate_grid}), which would then result in higher rates of weathering and removal of atmospheric \ch{CO2} (Figure~\ref{fig:schema}B, equations \ref{eq:fcarb}, \ref{eq:fsil}).
This would result in an equilibrium atmospheric \ch{CO2} that decreases with increasing luminosity in worlds that have continental weathering, represented by our simulations in the ML interior convection regime.\\

\noindent
Simulations in the stagnant-lid (SL) regime do not include weathering of emerged surfaces (continental or otherwise) but only seafloor weathering, which is also indirectly temperature-dependent (Figure~\ref{fig:schema}B, Equation \ref{eq:fdiss}).
Whether this would generally result in the lack of a relation between p\ch{CO2} and luminosity \citep{catling2018exoplanet}, or on the contrary in a significant effect of seafloor weathering on the regulation of atmospheric \ch{CO2} \citep[like it did on the early-Earth;][]{krissansen2017constraining}, is unresolved.\\

\begin{figure*}
    \centering
    \includegraphics[scale=.90]{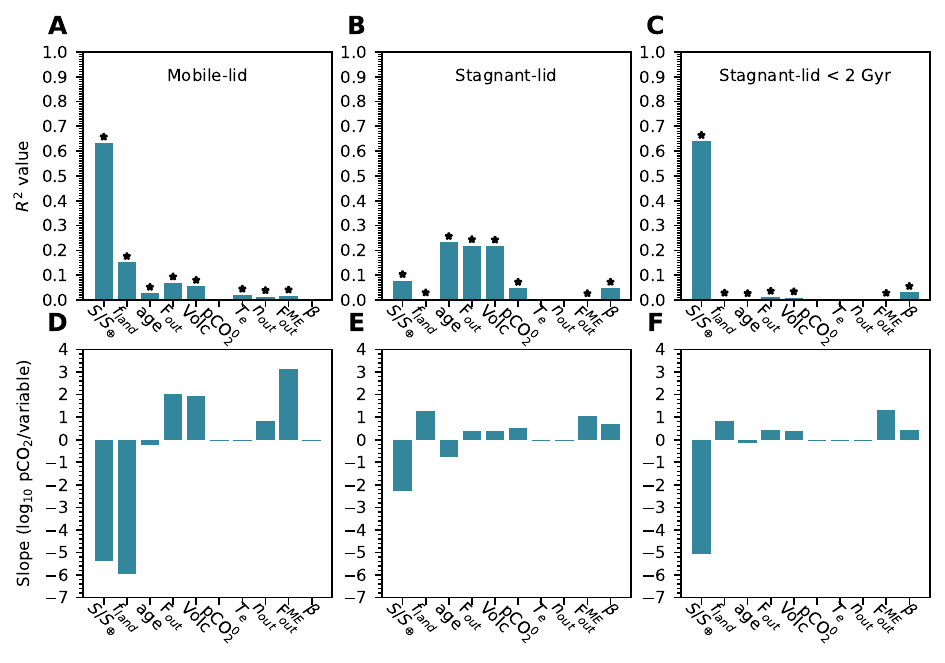}
    \caption{\textbf{Decomposition of linear correlation coefficients and inferred slope of different prior variables.}
    \textbf{A-C} Correlation coefficients for mobile-lid simulations (A), stagnant-lid simulations (B), and 'young' ($<2$~Gyr) stagnant-lid simulations (C).
    Stars indicate variables for which linear correlation is found at significance threshold $p<0.05$ (two-sided Wald test).
    \textbf{D-F} Inferred slopes of the different predictor variables for mobile-lid simulations (D), stagnant-lid simulations (E), and 'young' ($<2$~Gyrs) stagnant-lid planets.
    Variables not pictured were found to have p-values exceeding the $0.05$ significance threshold in all three cases \textit{or} to correlate at the $0.05$ significance level, but with $R^2<0.01$.
    Note that correlations with p\ch{CO2}$^0$, $F_{out}$, $F_{out}^{ME}$, and $Volc$ are performed on $\log_{10}$-transformed values of those variables.}
    \label{fig:decomposition}
\end{figure*}

\begin{figure*}
    \centering
    \includegraphics[width=\textwidth]{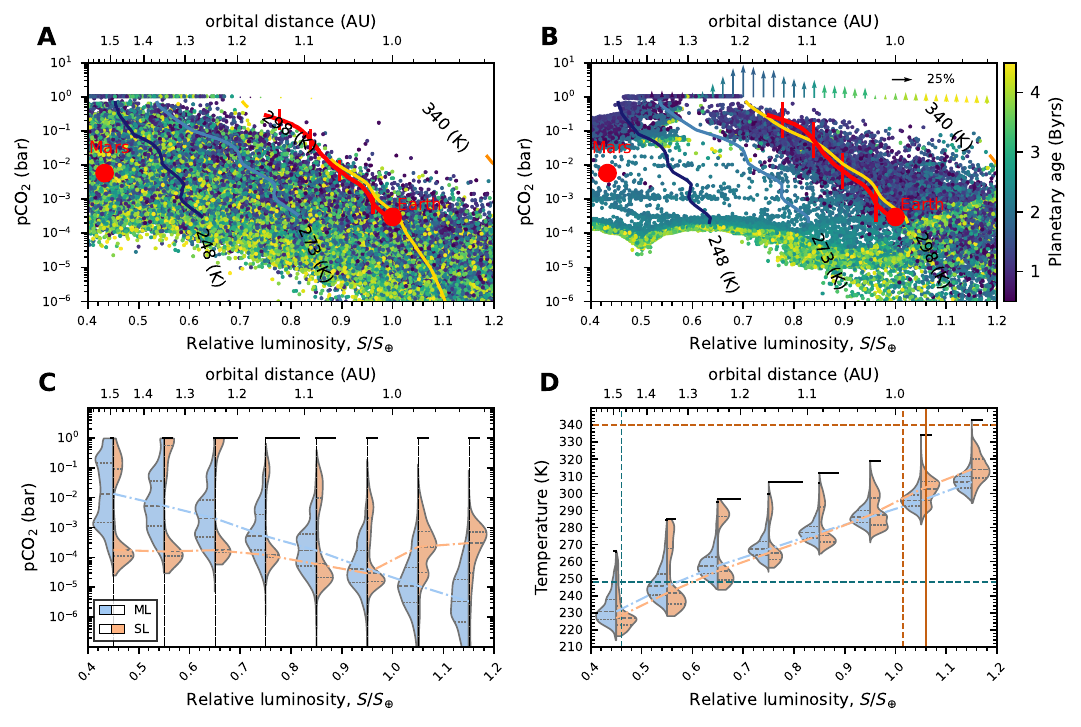}
    \caption{
    \textbf{Effect of relative luminosity, $S/\ssun$, on simulated steady-state atmospheres ($p\ch{CO2}$, temperature).} \textbf{A} Terrestrial exoplanets in the mobile-lid (ML) interior convection regime. \textbf{B} Terrestrial exoplanets in the stagnant-lid (SL) regime.
    Vertical arrows represent simulations that reach the maximal value of atmospheric \ch{CO2} of our grid with a positive derivative of p\ch{CO2}. Arrow size scales with the fraction of simulations in the binned interval that reach this \ch{CO2} limit.
    \textbf{C,D} Distributions of simulations across a discretized range of $S/\ssun$. In \textbf{A,B}, simulation points are colored according to the planetary age used as a model parameter (planets with an age similar to Earth appear in yellow). Red circles indicate modern Earth and Mars. Nominal Earth's trajectory \citep[change in luminosity and p\ch{CO2} over time;][]{krissansen2018constraining} is represented with the red line ending with modern values (red circle), intersecting vertical lines are placed at 1, 2, and 3~Byrs from formation. The dark blue, light blue, gold, and orange contour lines indicate temperatures of 248~K, 273~K, 298~K, and 340~K respectively, predicted from the climate model ran with fixed average values for $p\ch{CH4}$ and $p\ch{H2}$. Linear regression of $\log_{10}p\ch{CO2}$ versus $S/\ssun$ infers a slope (in units of $\log_{10}$(bar)~$\ssun^{-1}$) of $-5.37\pm 0.02$ (standard deviation) with correlation coefficient value $r^2=0.63$ in \textbf{A}, and $-2.26\pm 0.04$ with correlation coefficient value $r^2=0.08$ in \textbf{B}. 
    In \textbf{C,D}, blue (left part of each distribution graph) and orange (right) distributions are compiled from simulations in the ML and SL regimes, respectively. Dot-dashed lines indicate median values, and horizontal dark bars indicate simulations that have reached the \ch{CO2} modeling limit.
    In \textbf{D}, the horizontal blue dashed line (lower straight line) corresponds to the temperature of 248~K, at which \citet{charnay2013exploring} report global glaciation including at the equatorial band. The vertical blue dashed line (to the left) corresponds to the OHZ of 1-bar atmosphere planets reported by \citet{kopparapu2013habitable}. The horizontal orange dashed line (upper) corresponds to the 340~K temperature, at which \citet{kopparapu2013habitable} calculate that depletion of the water reservoir through hydrogen escape can occur on the timescale of Earth's age. The vertical orange dashed line (to the right) corresponds to the moist runaway inner Habitable Zone limit, and the solid orange line corresponds to the greenhouse runaway limit \citep{kopparapu2013habitable}.
    }
    \label{fig:trend_violins}
\end{figure*}

\noindent
Consistent with expectations, our simulations show a pronounced log-linear negative correlation between p\ch{CO2} and luminosity for ML terrestrial exoplanets (including `\ch{CO2}-accumulation' simulations), which have continental weathering (correlation coefficient $r^2 \approx 0.63$; Figure~\ref{fig:trend_violins}A).
The emerged land fraction $f_{land}$ is a notable driver of the variance in p\ch{CO2} for simulations of ML worlds (Figure~\ref{fig:decomposition}A), for which emerged land is assumed to be continental crust, contributing to "continental weathering" (Equation \ref{eq:fcont}).
Planetary age also negatively correlates with atmospheric \ch{CO2} (colored points on Figure~\ref{fig:trend_violins}A, and Figure~\ref{fig:decomposition}A), due to the relation that we impose between age and volcanic outgassing of \ch{CO2} (Equation \ref{eq:agefout_ML}).
Additional details on correlations between equilibrium p\ch{CO2} and simulation parameters can be found in Appendix \ref{sec:app:mreg}.\\

\noindent
In contrast, there is no clear correlation pattern of p\ch{CO2} and luminosity in simulations of SL planets (linear correlation coefficient $r^2 \approx 0.1$, Figure~\ref{fig:trend_violins}B, Figure~\ref{fig:decomposition}B).
However, `young' planets (blue dots in Figure~\ref{fig:trend_violins}B) have a stronger linear correlation between $\log_{10}$ p\ch{CO2} and $S/\ssun$ ($r^2=0.64$ among planets younger than 2~Gyr; Figure~\ref{fig:decomposition}C).
This puts luminosity as similar driver of atmospheric \ch{CO2} in both `young' SL and ML planets.
In addition, the inferred slope of this correlation is very close ($-5.04 \pm 0.03$~$\log_{10}$(bar)~$(S/\ssun)^{-1}$; Figure~\ref{fig:decomposition}F) to the luminosity-\ch{CO2} correlation slope estimated for ML simulations ($-5.37 \pm 0.02$~$\log_{10}$(bar)~$(S/\ssun)^{-1}$; Figure~\ref{fig:decomposition}D).\\

\noindent
In sum, despite not having continental weathering as a negative feedback to atmospheric \ch{CO2}, SL planets younger than 2~Gyrs may exhibit a correlation between \ch{pCO2} and luminosity similar to that of ML planets.
This suggests that samples of a few dozens of exoplanets may yield very similar correlation tests under the hypotheses that planets are in the SL are ML regimes.
This correlation is however lost with planet age, which in our model is a direct proxy for volcanic outgassing (Figure~\ref{fig:schema}C).
Through Equation (\ref{eq:fdiss}), the seafloor weathering feedback to atmospheric \ch{CO2} depends on the rate of volcanic outgassing, as fresh basalt need to be resupplied for weathering to occur continuously.
Hence, SL planets may lack long-term control of atmospheric \ch{CO2}, not because they lack continental weathering, but because their volcanic output may halt earlier than that of ML planets.\\

\noindent
In our simulations, older SL planets (age $>2$~Gyr) show a more complex distribution of p\ch{CO2}.
At relatively low luminosities ($S/\ssun<0.7$), these simulations cluster at relatively low \ch{CO2} abundances (Figure~\ref{fig:trend_violins}B).
The simulations of these older SL planets are characterized by low or non existent weathering and volcanic dynamics (Figure~\ref{fig:schema}C).
As a result, their atmospheric \ch{CO2} dynamics should be dominated by photochemistry.
The analysis of the balance between photochemical sources and sinks in those simulations supports that.
Indeed, in calculations in which p\ch{CH4} and p\ch{H2} are constrained to their simulation-average final value, net \ch{CO2} photochemical production is predicted when p$\ch{CO2}\lessapprox 10^{-4}$~bar and net removal when p$\ch{CO2}\gtrapprox 10^{-4}$~bar (Figure~\ref{fig:photochem_fluxes}).
Hence, under this approximation on p\ch{CH4} and p\ch{H2}, p$\ch{CO2}\approx 10^{-4}$~bar is predicted to be the steady state value of atmospheric \ch{CO2} based solely on photochemical fluxes.
Under the alternate approximation of p\ch{CH4} and p\ch{H2} being fixed at their average final value in simulations that have final p\ch{CO2}$>10^{-1}$~bar, photochemical calculations show that an unstable point exists for \ch{CO2} at $S/\ssun>0.7$, allowing \ch{CO2} to potentially accumulate above 1~bar (Figure~\ref{fig:photochem_fluxes}B).

\noindent
Analysis of photochemical dynamics of \ch{CO2} (Figure~\ref{fig:photochem_fluxes}B) shows that simulations of SL planets older than 2~Gyrs may exhibit bistability with respect to their steady-state atmospheric \ch{CO2}.
The specifics dynamics of p\ch{CO2} in these cases are dependent upon the dynamics of other atmospheric gases included in the network of photochemical reactions.
This non-trivial dynamic behavior of atmospheric \ch{CO2} allows the initial value of p\ch{CO2} in SL simulations with age$>2$~Gyrs to be a significant driver of the simulation's equilibrium p\ch{CO2} (unlike in ML simulations; see Figures~\ref{fig:decomposition}A-C).
In consequence, contingent events may play an important role in the atmospheric evolution of planets in the SL interior convection regime that are older than 2~Gyrs.
For instance, an individual volcanic event occurring 2 billion years or more after planetary formation could result in an abrupt change in steady-state p$\ch{CO2}$.
Hence, there might be fundamental unpredictability of the atmospheric concentration of \ch{CO2} for SL planets that have exhausted their internal heat budget.

\subsection{Influence of interior convection regime on climate and habitability.}
Most (about 80\%) simulations in both SL and ML interior convection regimes have temperatures comprised between 248~K \citep[global glaciation;][]{charnay2013exploring} and 340~K \citep[water runaway;][]{kopparapu2013habitable}.
ML simulations (with continental weathering) are very slightly more likely to have a temperate climate (79.9\%) than their SL counterparts without continental weathering (79.3\%).
Glacial climates ($T_{surf}<248$~K) were virtualy as frequent in both regimes (20.1\%).
Runaway hot climates ($T_{surf}>340$~K) were close to equally rare in SL simulations (0.6\%) and in ML simulations (0\%).\\

\noindent
In spite of fundamental unpredictability of p\ch{CO2} on SL planets, their climate turns out to be rather narrowly constrained, and similar to ML planets (Figure~\ref{fig:trend_violins}D).
This challenges the view that SL planets would in general have unstable climates and experience either snowball or greenhouse runaway.
Instead, terrestrial planets in the SL regime could still be frequently observed in a temperate climate stage of their geological history ($T_{surf} \in [248,340]$~K), despite a relatively short-lived volcanic output.
This can be explained by photochemistry providing a stabilizing effect on p\ch{CO2}, (Figure~\ref{fig:photochem_fluxes}B).
Although photochemistry has a weak effect on p\ch{CO2}, with little dependence on luminosity, the stabilization outcome observed at low volcanic outgassing (or equivalently with age $>2$~Gyrs) may result --by chance-- in a temperate climate.\\

\noindent
Noticingly, at low values of relative luminosity ($S/\ssun<0.7$), a subset of simulations in the SL regime lead to more clement temperature ($\approx 280$~K) than their ML counterparts (Figure~\ref{fig:trend_violins}D).
SL simulations often have higher equilibrium p\ch{CO2} than ML simulations at these luminosities (Figure~\ref{fig:trend_violins}C).
Therefore, SL planets may be more likely to sustain liquid water on their surface at low luminosity/large orbital radius than ML planets would.

\subsection{Predictions at planet population scale and prospects for future testing of hypotheses}
Our results show that terrestrial planets in the SL regime may not be statistically less frequently habitable than planets in the ML regime.
Yet, this may largely be unrelated to the odds of \textit{inhabitation} of planets in the ML vs. SL regimes, which depends on unknown or poorly constrained factors.
These odds could actually differ markedly.
It may be that when in the SL regime, terrestrial planets have a climate more sensitive to perturbations, as hinted by the bimodal distribution shown in Figure~\ref{fig:trend_violins}B (and Figure~\ref{fig:photochem_fluxes}B).
Thus, advancing the search for biological activity on terrestrial exoplanets would still benefit from observations designed to test whether a sample of terrestrial planets in the HZ of their host star are in the ML or SL regime.\\
\noindent
To assess the performance of different observation strategies in inferring the most plausible dominant interior convection regime in terrestrial exoplanets, we calculate the risk of insufficient Bayesian evidence in favor of whichever hypothesis we assume is correct.
Hypotheses to be tested using a sample of exoplanet atmosphere data are defined as: 
\begin{hypothesis}{$\mathrm{H}_{ML}$: }
All terrestrial planets in the HZ have both continental weathering and seafloor weathering (ML regime).
\end{hypothesis}
\begin{hypothesis}{$\mathrm{H}_{SL}$: }
All terrestrial planets in the HZ only have seafloor weathering (SL regime).
\end{hypothesis}
To prospectively test these hypotheses, we focus on $p\ch{CO2}$ as an observable, and the incident flux $S/\ssun$ as a 'contextual' observable \citep{catling2018exoplanet}.\\

\noindent
To measure the performance of an observation strategy (defined below), we calculate the probability or risk that a sample $X$ resulting from this strategy yields Bayesian evidence in favor of the correct hypothesis $\text{BF}_{\mathrm{H}_{true},\mathrm{H}_{false}}$ smaller than $\eta$ :
\begin{equation}
    \label{eq:inferencerisk}
    \begin{split}
        \rho = max [ & P(\text{BF}_{\mathrm{H}_{ML},\mathrm{H}_{SL}}(X^{\mathrm{H}_{ML}})<\eta) ,\\
        & P(\text{BF}_{\mathrm{H}_{SL},\mathrm{H}_{ML}}(X^{\mathrm{H}_{SL}})<\eta)] \space .
    \end{split}
\end{equation}
Bayesian evidence is calculated using the Bayes factor $\text{BF}$ \citep[see \textit{e.g.}][]{kass1995bayes}
\begin{equation}
    \label{eq:bayesratio}
    \text{BF}_{\mathrm{H}_a,\mathrm{H}_b} = \frac{P(X|\mathrm{H}_a)}{P(X|\mathrm{H}_b)} \space .
\end{equation}
Equation (\ref{eq:bayesratio}) quantifies how much more likely the sample $X$ is under $\mathrm{H}_a$ than under $\mathrm{H}_b$.
{While $\text{BF}_{\mathrm{H}_a,\mathrm{H}_b}>1$ technically constitutes evidence in favor of $\mathrm{H}_a$, it is common to classify Bayesian evidence according to Jeffrey's scale: \emph{anecdotal} ($\text{BF}_{\mathrm{H}_a,\mathrm{H}_b} \in [1,3[$), \emph{substantial} ($[3,10[$), \emph{strong} ($[10,100[$), and \emph{very strong} or \emph{decisive} \citep[$\geq 100$;][]{kass1995bayes,jeffreys1967theory}.}\\

\begin{figure*}
    \centering
    \includegraphics[scale=1]{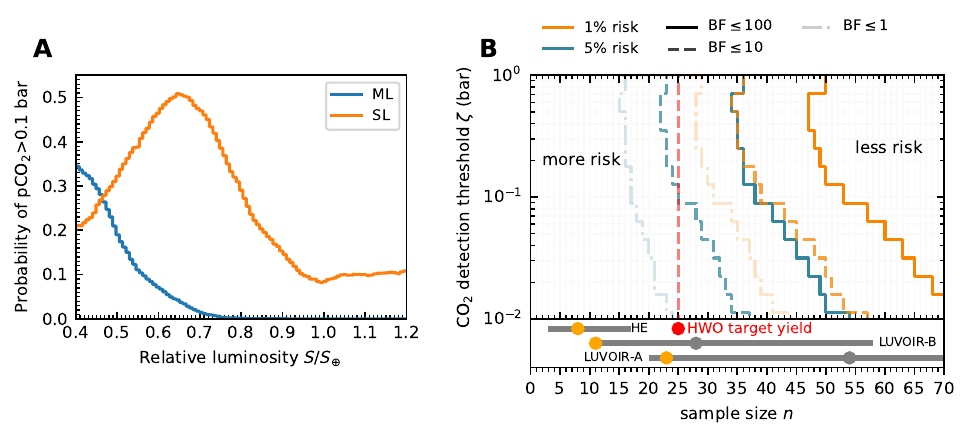}
    \caption{\textbf{Probabilistic calculations for the inference of the dominant interior convection regime.}
    \textbf{A.} Estimate of the conditional probability $P(\Gamma|S/S_{\oplus})$ for each interior convection scenario used in equation \ref{eq:proba_estimate}.
    \textbf{B}. Inference risk (Equation \ref{eq:inferencerisk}) over strategies of \ch{CO2} detection threshold ($\zeta$) and sample size ($n$).
    Orange (blue) lines indicate  1\% (5\%) risk contour levels. Continuous lines map the risk that evidence in favor of the correct hypothesis is less than ``extreme" (BF$<$100). Dashed lines are contours of the risk that evidence in favor of the correct hypothesis is less than ``strong" (BF$<$10). Semi transparent dashed and dotted lines are contours of the risk that a sample yields stronger evidence in favor of the incorrect hypothesis than in favor of the correct one.
    The inset at the bottom shows the expected yield of different concepts of space observatories: Large UV/Optical/IR Surveyor \citep[LUVOIR,][]{luvoir2019luvoir}, HE: Habitable Exoplanet Observatory \citep[HabEx,][]{gaudi2020habitable}, and Habitable Worlds Observatory \citep[HWO,][]{mamajek2023}.
    The gray lines and circles indicate the expected number of exo-Earth candidates (EECs) detected by each mission.
    Orange dots indicate the expected number of EECs spectra from each mission.
    The red dot indicates the target EEC spectra yield from the most recently proposed mission concept HWO.}
    \label{fig:riskmap}
\end{figure*}

\noindent
In Equation (\ref{eq:inferencerisk}), the Bayes factor is calculated in the case where $\mathrm{H}_{SL}$ is true and $\mathrm{H}_{ML}$ is the \emph{null hypothesis}, as well as in the case where $\mathrm{H}_{ML}$ is true and $\mathrm{H}_{SL}$ is the \emph{null hypothesis}.
We define an observation strategy as the combination of the number of observed targets ('sample size', $n$), and the minimal abundance of atmospheric \ch{CO2} that can be detected by the hypothetical instrument ('\ch{CO2} detection threshold', $\zeta$).
The resulting sample is a series of pairs $X_{n,\zeta}=\{(S/\ssun)_i,\Gamma_i\}_{i \in [1,n]}$, where $\Gamma_i = 1$ if $p\ch{CO2}|_i>\zeta$ and 0 otherwise.
The likelihood of such a sample under hypothesis $\mathrm{H}$ is then expressed as
\begin{equation}
    \begin{array}{rcl}
        P(X_n | \mathrm{H}) & = & P(\{\Gamma_i,(S/\ssun)_i\}_{i \in [1,n]}|\mathrm{H}) \\
         & = & \prod_{i\in [1,n]} P((\Gamma_i,(S/\ssun)_i)|\mathrm{H})\space .
    \end{array}
\end{equation}
We approximate the values of $P((\Gamma_i,(S/\ssun)_i)|\mathrm{H})$ using our simulations by discretizing $S/\ssun$ and assessing the frequency of $\Gamma$ in the resulting intervals
\begin{equation}
    \label{eq:proba_estimate}
    P((\Gamma_i,(S/\ssun)_i)|\mathrm{H}) \approx \frac{\sum_{j \in \mathbf{k}_{\mathrm{H}}}\Gamma(p\ch{CO2}|_j)}{Card(\mathbf{k}_{\mathrm{H}})}
\end{equation}
where the vector $\mathbf{k}_{\mathrm{H}}$ contains $Card(\mathbf{k}_{\mathrm{H}})$ indices $j$ of all simulations of hypothesis $\mathrm{H}$ that verify $(S/\ssun)_j \in [(S/\ssun)_i- 0.05,(S/\ssun)_i + 0.05[$ as illustrated by Figure \ref{fig:riskmap}A.\\

\noindent
The number of possible combinations of $\{\Gamma_i,(S/\ssun)_i \}_{i \in [1,n]}$ increases rapidly with the sample size $n$ (for instance, positive \ch{CO2} detection in 15 candidates out of 30 corresponds to $10^{20}$ sets of pairs of $(\Gamma,S/\ssun)$).
Thus, approximating $P(\text{BF}(X)>\eta)$ requires to resample a large number of samples $X$ from the simulated populations of SL and PT.
We find that resampling 200,000 times is largely sufficient to estimate $P(\text{BF}(X)>\eta)$ from the frequency of $\text{BF}(X)>\eta$, even at relatively large sample size.\\

\noindent
The probability that a survey results in insufficient evidence decreases with sample size $n$ (Figure~\ref{fig:riskmap}B).
Surprisingly, our analyses show that \ch{CO2} detection thresholds $\zeta$ lower than $\sim 0.1$~bar do not improve the odds of correctly inferring the dominant interior convection regime (Figure~\ref{fig:riskmap}B).
In our hypothetical samples, characterization of atmospheric \ch{CO2} is assigned a boolean variable according to whether the actual \ch{CO2} partial pressure is greater than $\zeta$ or not.
Hence, values of $\zeta$ that are either too high or too low may completely eliminate the possibility to discriminate between the SL and ML distributions.
Our analyses show that unless accompanied with means of constraining the absolute value of p\ch{CO2}, detection threshold values lower than 0.1~bar in fact reduce the expected reduce ability to discriminate between regimes.\\

\noindent
For $\zeta>0.1$~bar, lower than 1\% risk of achieving less than very strong evidence ($\text{BF}<100$) is reached for a sample size exceeding 47-50 planets (Figure~\ref{fig:riskmap}B).
Strong evidence (BF$>10$) can be achieved in $\geq 99\%$ of samples with a sample size as small as 35.
The same strategy would achieve very strong evidence (BF$>100$) in 95\% of samples (solid blue line in Figure~\ref{fig:riskmap}B).
Hence, such a strategy ($n>35$) could be optimal if 5\% risk of achieving less than decisive evidence and 1\% risk of achieving less than strong evidence is acceptable.\\

\noindent
The target yield for exo-Earths candidate planetary spectra of LUVOIR \citep[11--23,][]{luvoir2019luvoir} and HabEx \citep[8,][]{gaudi2020habitable} could lead to relatively high risk of incorrect inference (Figure~\ref{fig:riskmap}B, inset, orange dots).
The target planetary spectra yield for the Habitable Worlds Observatory \citep[25,][]{mamajek2023} is projected to yield strong evidence in $>95$\% of samples, but decisive evidence in only $80-85$\% of samples.
However, these probabilities are conditional on the instrument being able to produce acceptable signal to noise ratio for a \ch{CO2} spectral feature corresponding to \ch{CO2} less abundant than 1~bar in the atmosphere of \emph{all} characterized candidates.

\section{Discussion}

In our simulations, the interior convection regime (stagnant or mobile lid) only had a minor influence on the expected frequency of terrestrial planets with a temperate climate among those that are in the Habitable Zone (HZ) of their host star; the frequency is only slightly lower if stagnant-lid (SL), rather than mobile-lid (ML), is the typical interior convection regime.
A strength of our study is the large number of planets for which we have been able to simulate their climate evolution, accounting for different geochemical mechanisms that we have coupled together.
As a trade-off of building such a coupled and general model, we have had to make simplifying assumptions, leading  to several caveats in how our result can be used to tackle future observations of the atmospheres of terrestrial exoplanets.
We used a 1-D climate model, and interpolated between pre-simulated grid points which only gives a rough approximation of the surface temperature (Figure \ref{fig:clima_comparisons}).
We also initialized these climate simulations with a vertical water profile that is not the one typically used in calculations of the inner edge of the HZ (greenhouse runaway).
All of our simulations assume that surface liquid water is present, and 2-4\% of our simulations fail because they dynamically exit our pre-run photochemical/climate grid such that fitting them into these analyses remains difficult.
\\

\noindent
Despite that, we did find simulations that have final states inconsistent with the habitable zone assumptions.
Some simulations rest at surface temperature exceeding the estimated onset of greenhouse runaway \citep[340~K;][]{kopparapu2013habitable} or below the threshold for global freezing \citep[248~K][]{charnay2013exploring}.
As described in the Results section, the frequency of these occurrences is virtually the same between scenarios.
Because we use a constant vertical water profile in our simulations, and limit ourselves to atmospheric states that permit radiative equilibrium while being consistent with surface liquid water, it is expected that our simulations underestimate the frequency of greenhouse runaway.
\\

\noindent
Hence, our work cannot conclusively quantify the frequency of habitable surface conditions that one may expect in terrestrial planets in the HZ.
Instead, our simulations suggest that in a sample of terrestrial planets in the HZ, whichever interior convection regime is most common should not play a critical role in determining the fraction of these planets that have surface conditions permitting liquid water at a given observation time.
However, our simulations align with expectations that interior convection regime may be critical in determining whether a terrestrial planet can sustain surface liquid water over more than $\approx 2$~Gyrs.
As a consequence, it is useful to design observational strategies that may enable a future telescope to infer which end-member interior convection regime terrestrial planets orbiting other FGK stars lean more towards.
\\

\noindent
In samples drawn from our simulations, inference of the interior convection regime that dominates in terrestrial exoplanets in the HZ is achieved with ``decisive evidence'' ($\mathrm{BF}\geq 100$) in $\geq 99\%$ of samples of 50 exoplanets or more for which detection of \ch{CO2} is guaranteed if at a partial pressure of $0.1$~bar or above.
Inference in a sample size of 25 yields strong true positive evidence ($\geq 10$) in $\geq 95\%$ of samples.\\

\noindent
Bayes factor approaches, such as the one we use, quantify the information gain of sampling, making our analysis agnostic to prior beliefs on which convection regime is most frequent \citep{jeffreys1967theory}.
Our simulations assume a random initial state and parameter values, and are solved using fixed values for age-dependent variables (e.g. volcanic outgassing).
Thus the results are robust to uncertainties regarding stochastic events that could influence the global climate such as impacts or individual volcanic events, as such events can be assumed time-independent on a billion-year timescale.
Additionally, because we solve for steady-states, our simulations do not necessarily assume that any given planet remains in a specific interior convection regime across geological timescales; but rather that the atmosphere would move from a steady-state to another.
\\

\noindent
Crafting a null hypothesis is crucial to interpret empirical observations.
It is not obvious what should be the null hypothesis whose rejection would mean that terrestrial planets have continental and seafloor weathering as a mechanism that ensures temperate climates across the range of luminosity that spans the HZ.
Examples in the solar system \citep{solomatov1996stagnant}, as well as recent planetary interior modeling \citep{dorn2018outgassing,foley2018carbon} allowed us to flesh out a credible alternate hypothesis.
Namely, we leveraged the possibility that terrestrial planets may have a weaker mode of silicate-carbonate weathering, as a plausible alternative to the hypothesis that planetary habitability within the HZ is supported by Earth-like weathering of sub-aerial continental crust in addition to seafloor dissolution.\\

\noindent
Despite geophysical control over atmospheric \ch{CO2} existing in both interior convection regimes, the surface temperature is found to increase with luminosity.
Thus, the direct effect of increasing luminosity on surface temperature overcompensates the indirect effect of decreasing the equilibrium p\ch{CO2}.
Barring episodes of snowball or hothouse Earth, others who used a model slightly simpler than the one presented here, have suggested that on the billion-year timescale, there has been no or very little trend in the Earth's global surface temperature \citep{krissansen2018constraining}.
Our simulations do not predict this outcome as likely under a single interior regime convection Figure~\ref{fig:trend_violins}A,B).
It is generally assumed that Earth's climate was stabilized by the carbonate-silicate geological carbon cycle (as well as seafloor weathering) reducing the concentration of \ch{CO2} in the atmosphere as the luminosity of the Sun increased.
By relaxing the relation between age and luminosity assumed for calculations of Earth's climate evolution it appears that seafloor weathering neither alone nor in combination with continental weathering is a sufficient control of atmospheric carbon to entirely compensate the effect of luminosity.
As a consequence, it is possible that the Earth's relatively stable temperate climate over geological times is not due entirely to the stabilizing effect of the carbonate/silicate weathering cycle, but also to the concurrent changes of the Sun's luminosity (increasing) and of the Earth's outgassing rate (decreasing), as well as the timing of continental emergence.\\

\noindent
\citet{Tosi2017} and \citet{foley2018carbon} have suggested that the stabilization of the climate of SL planets may be limited by the supply of fresh rock associated with volcanic outgassing.
In agreement with these expectations, using the outgassing parameterization proposed by \citet{krissansen2017constraining}, together with the age-outgassing parameterization of \citet{dorn2018outgassing} results in simulations of SL regime planets that show almost no weathering climate feedback after $\approx 2$~Byrs.
This further confirms that planetary age plays a critical role in the habitability of SL planets \citep{foley2018carbon}.\\

\noindent
While parameters of the carbon-cycle model and for the age-outgassing relation for the Earth are sampled in ranges grounded in laboratory studies \citep[][; Appendix Table~\ref{table:parameters}]{krissansen2017constraining}, the availability of alternate models for the evolution of volcanic outgassing on SL worlds is limited \citep[an alternative to the parameterization that we used can be found in][]{foley2018carbon}, and their uncertainty is poorly evaluated.
Further research of the geological past of Mars and Venus is required to ground models of the evolution of the outgassing rate of stagnant lid worlds and constrain its variability.\\

\noindent
The Earth, Venus, and perhaps Mars, have not been in any one single tectonic regime during the entirety of their geological history, instead they may have undergone shifts between modes, a regime described as 'episodic' \citep{moresi1998}.
This scenario is not explicitly modeled here.
By running our model from random initial conditions until an equilibrium for a fixed outgassing rate is found, each simulation is agnostic to the planet's past.
However, the planet's internal heat dissipates at different rates according to their mantle's convection regime.
Hence, if a planet changes convection regime after significant time spent in the alternate one, their outgassing rate at a given age will likely differ significantly from the parameterizations we use here.
Therefore, the luminosity-p\ch{CO2} distributions that we simulate under the two end-member regimes should not be affected by planets having undergone a different tectonic mode in their past, but the age-p\ch{CO2} distributions might.\\

\noindent
Our simulations are seeded with random initial conditions, thus assessing the possibility for multiple equilibria to exist for a given pair of values for outgassing and luminosity.
Doing so reveals that the absence of continental weathering and outgassing on SL planets older than 2~Byrs may cause bistability of their climate and atmospheric composition.
An event releasing large amounts of \ch{CO2} to the atmosphere, without equivalent supply of fresh rock, would then result in an inescapably \ch{CO2}-rich atmosphere, regardless of luminosity.
However, our model also suggests that such random events may not be required for \ch{CO2} runaway atmospheres under relatively high incident luminosity, as photochemistry can then lead to slow \ch{CO2} accumulation (Figure~\ref{fig:photochem_fluxes}B).\\

\noindent
Our model revealed that weak or absent outgassing, limited or absent supply of weatherable rock (as one might expect on a relatively old world subject to stagnant-lid mantle convection), leads to atmospheric \ch{CO2} dynamics being dominated by photochemistry, resulting in equilibrium p\ch{CO2} (high or low) dependent on simulation initial conditions when luminosity is between 0.7 and 1~$\ssun$.
Hence, if certain trajectories of atmospheric evolution on the billion years scale favor one scenario over the other  (\ch{CO2}-rich or \ch{CO2}-poor), then warm climates on SL worlds may be more frequent than we predict (if \ch{CO2}-rich is more likely), or the overlap in the luminosity-\ch{CO2} distribution between SL and ML worlds may be greater than we estimate (if \ch{CO2}-poor is more likely).
Albeit beyond the scope of this study, the coupled model presented here could be used to run long-term (several billion years) simulations, and include random perturbations in order to better constrain the long-term climate dynamics.
This could then also be used to assess the resilience of the climate and atmospheric state of planets in different tectonic scenarios.\\

\noindent
The sample size targeted by the Habitable Worlds Observatory (HWO) is 25 Earth-mass planets in the HZ of FGK stars \citep{mamajek2023}.
\citet{lehmer2020carbonate} calculated that such sample size would be insufficient to reject log-uniform distribution of \ch{CO2} with respect to luminosity.
By simulating atmospheric \ch{CO2} under an explicit hypothesis alternate to Earth-like geological carbon cycle, informed by our knowledge of Earth's past and other terrestrial planets in the Solar system, we showed that correlation tests may fail to discriminate between interior convection regimes due to young SL planets.
However, the joint distributions of p\ch{CO2} and luminosity between regimes differ markedly (Figure \ref{fig:riskmap}A), allowing likelihood ratio approaches to perform well at sample sizes realistic for future space telescopes (Figure \ref{fig:riskmap}B).
Specifically, we find that a sample size of 25 is likely ($\geq 80$\%) to provide decisive evidence ($>100$), and extremely likely ($\approx 96$\%) to provide strong evidence ($>10$) for the inference of the dominant interior regime.
According to our model, this evidence yield can be achieved for a sample size of 25 exoplanets if the characterization effort for each of them guarantees that \ch{CO2} above $0.1$~bar would lead to positive detection in the spectrum.
However, the proposed target sample for the HWO includes different star types while our model assumes a Sun-like (type G) star.
If variation in star spectra across the HWO target sample was to significantly alter the distribution of atmospheric \ch{CO2}, then the required sample size could differ from our estimate.
In addition, whether a sample size of 25 roughly Earth-mass planets orbiting FGK stars is attainable within the constrains posed on future instruments is uncertain.
Earth-like planets around FGK stars have not been yet detected, possibly due to instrument limitations, leading yield estimates of missions to rely on occurrence rates of such planets (noted $\eta_{\bigoplus}$) obtained from models.
The PLATO mission \citep{PLATO2024}, scheduled to launch in 2026 is designed to fill this gap, and permit the detection and characterization of bulk properties of Earth-sized planets orbiting FGK stars.\\

\noindent
The distribution of the radius of planets in the target sample of the HWO varies around one Earth radius.
Planetary radius (more precisely planetary mass) is a key parameter determining the heat budget and thus affects not only the volcanic outgassing rate but also the likelihood of different interior convection regimes \citep{dorn2018outgassing}.
As a consequence, the population from which the Habitable Worlds Observatory would sample could differ from the one we simulate.
In particular, by adding variability in planetary radius, it is likely that the spread of the expected $S$-p\ch{CO2} relation be increased, thus increasing the overlap between the two scenarios of interior convection regimes and subsequently increasing the required sample size.\\

\noindent
The required sample size might, however, be reduced, in three ways.
The prospective analysis presented here assumes that the sample is composed of independent observations, however, one may increase efficiency by adopting a sequential strategy such that each new observation is targeted in such a way that its addition to a known sample maximizes information gain.
With moderate effort, one could refine the fixed Bayes Factor design we have studied into a so-called Sequential Bayes Factor design and analyze its effectiveness \citep{schonbrodt2017bayes}.
\citet{bixel2020testing} have suggested using inferred planetary age to increase the statistical power of life detection surveys.
Apart from the interior convection regime, atmospheric abundance of \ch{CO2} may hence be chiefly determined by luminosity/orbital radius, planetary age, and planetary mass.
All of these quantities could be estimated when characterizing an exoplanet.
Planetary mass and age could eventually be included in a more sophisticated prospective statistical test in order to quantify the gain of retrieving such parameters on the required sample size.
Future work may also aim at relaxing the assumption of 1 Earth-mass planets by linking volcanic outgassing to planetary mass \citep[\textit{e.g.} following][]{dorn2018outgassing}, as well as the assumption of a strictly Sun-like star by running additional simulations of the atmospheric model \textit{Atmos}.\\

\noindent
Ultimately, the feasibility of the 0.1 bar limit of detection of \ch{CO2} depends on exoplanet survey mission design on the one hand, and on the distance of the sampled exoplanets to us and to their star in the case of a direct imaging survey such as the Habitable Worlds Observatory, or on the orbital period for transit-based proposed architectures \citep[\textit{e.g.} the \textit{Nautilus} concept;][]{apai2019nautilus}.
Whether such limit of detection can be achieved for 25 candidates will be critically important for the design of any future exoplanet observatory.
In future work, coupling our simulations with planetary spectra generators \citep{Villanueva2018} and survey simulators that integrate exoplanet catalog to simulate realistic exoplanet samples \citep{bixel2021bioverse} will help link instrument design parameters with our \ch{CO2} limit of detection parameter.
Doing so will enable to more directly identify the optimal compromise between instrument cost and scientific gain.
\section*{acknowledgments}
We are grateful for discussions with Alex Bixel, and members of the Alien Earths program supported by the National Aeronautics and Space Administration under Agreement No. 80NSSC21K0593) and the Nexus for Exoplanet System Science (NExSS) research coordination network sponsored by NASA’s Science Mission Directorate. 
AA, BS, and RF acknowledge support from France Investissements d’Avenir programme (grant numbers ANR-10-LABX-54 MemoLife and
ANR-10-IDEX-0001-02 PSL) through PSL–University
of Arizona Mobility Program, and from the US National Science Foundation, Dimensions of Biodiversity (DEB-1831493), Biology Integration Institute-Implementation (DBI-2022070), Growing Convergence in Research (OIA-2121155), and
National Research Traineeship (DGE-2022055) programmes.
%
\vspace{2mm}
\facilities{\textit{BIOCLUST} HPC Institut de Biologie de l'École Normale Supérieure in Paris France}
\software{
    \textit{Atmos}, Arney et al. 2016;
    SciPy, Virtanen et al. 2020
    }

\newpage
\appendix
\restartappendixnumbering
\label{appendix}

\section{Model parameters}

{
\defcitealias{krissansen2017constraining}{K-T\&C 2017}
\defcitealias{sauterey2020}{Sauterey+ 2020}
\defcitealias{kharecha2005coupled}{Kharecha+ 2005}
\begin{table}[ht]

    {
    \scriptsize
    
    \centering
    \begin{tabular}{|c c c c c c|}
    \hline
         parameter    & meaning & distribution & range & unit & reference\\ \hline
         $p\ch{H2}^0$  & initial $p\ch{H2}$  & Log-Uniform & $10^{-5}-10^{-2}$& bar & \\
         $p\ch{CO2}^0$ & initial $p\ch{CO2}$ & Log-Uniform & $10^{-4}-10^{-2}$& bar & \\
         $p\ch{CH4}^0$ & initial $p\ch{CH4}$ & Log-Uniform & $10^{-6}-10^{-4}$ & bar & \\
         $p\ch{CO}^0$  & initial $p\ch{CO}$  & Log-Uniform & $10^{-6}-10^{-4}$ & bar & \\
         \ch{pH^0}   & initial \ch{pH}   & Uniform     & 6.5-7.5 &  & \\
         $f_{land}$    & emerged land surface fraction  & Uniform & 0.0-0.35 & & \\
         $\tau$        & planet age  & Uniform & 0.5-4.5 & Gyr & \\
         $S/S_{\oplus}$     & Relative luminosity  & Uniform & 0.4-1.2 & $S_{\oplus}$ & \\
         $E_{bas}$ & Activation energy of basalt dissolution & Uniform & 60-100  &kJ~mol$^{-1}$ & \citetalias{krissansen2017constraining}\\
         $F_{out}^{ME} \times S_p$ & Present day volcanic outgassing of $\ch{CO2}$  & Uniform & $6-10\times 10^{12}$  & mol~yr$^{-1}$ & \citetalias{krissansen2017constraining}\\
         $\chi$ & $\ch{CO2}$ parameter for $F_{carb}$ & Uniform & 0.1-0.5 & & \citetalias{krissansen2017constraining} \\
         $\alpha_s$ & $\ch{CO2}$ parameter for $F_{sil}$ & Uniform & 0.1-0.5  & &  \citetalias{krissansen2017constraining} \\
         $\beta$ & Basalt dissolution sensitivity to basalt supply & Uniform & 0-2  & & \citetalias{krissansen2017constraining}\\
         $\gamma$ & Basalt dissolution sensitivity to pH & Uniform & 0.0-0.5 & & \citetalias{krissansen2017constraining}\\
         $T_{e}$ & weathering e-folding temperature & Uniform & 10-40 & K & \citetalias{krissansen2017constraining}\\
         $F_{carb}^{ME} \times S_p$ & Carbonate weathering flux on modern Earth & Uniform & $0.7-1.7 \times 10^{12}$  & mol~yr$^{-1}$ & \citetalias{krissansen2017constraining}\\
         $F_{sil}^{ME} \times S_p$ & Silicate weathering flux on modern Earth & Uniform & $0.7-1.7 \times 10^{12}$  & mol~yr$^{-1}$ & \citetalias{krissansen2017constraining}\\
         $F_{diss}^{ME} \times S_p$ & Basalt dissolution rate on modern Earth & Uniform & $0.2-0.9\times 10^{12}$ & mol~yr$^{-1}$ & \citetalias{krissansen2017constraining}\\
         $a_{grad}$ & Deep ocean temperature gradient parameter & Uniform & 0.75-1.4 & & \citetalias{krissansen2017constraining} \\
         $n_{carb}$ & Carbonate precipitation sensitivity to saturation & Uniform & 1.0-2.5 & & \citetalias{krissansen2017constraining}\\
         $n_{out}$ & Spreading-rate to age power law exponent & Uniform & 0.0-0.5 &  & \citetalias{krissansen2017constraining}\\
         $m$ & Spreading-rate to outgassing power law exponent & Uniform & 1.0-2.0 &  & \citetalias{krissansen2017constraining}\\
    \hline
    \end{tabular}
    \caption{Planetary parameters and initial values of time-dependent variables and their distributions. }
    \label{table:initialization}
    }
\end{table}}

\newpage
{\defcitealias{krissansen2017constraining}{K-T\&C 2017}
\defcitealias{sauterey2020}{Sauterey+ 2020}
\defcitealias{kharecha2005coupled}{Kharecha+ 2005}
\begin{deluxetable}{l c c l l}[H]
\caption{\textbf{Physical and chemical constants.} $^{\dagger}$ set to match the range for the modern value of $F_{diss}$ in \citet{krissansen2017constraining}, see Appendix Table~\ref{table:initialization}.}
\label{table:parameters}
\tablehead{\colhead{Parameter} & \colhead{Value} & \colhead{Unit} & \colhead{Description} & \colhead{Reference}}
\startdata
			$v_{\ch{H2}}$  & $1.3\times10^{-2}$  &$\text{cm}~\text{s}^{-1}$ & SBL piston velocity of $\ch{H2}$ &\citetalias{sauterey2020} \\
			$v_{\ch{CO2}}$ & $4.8\times10^{-3}$  &$\text{cm}~\text{s}^{-1}$ & SBL piston velocity of $\ch{CO2}$ &\citetalias{sauterey2020} \\
			$v_{\ch{CH4}}$ & $4.5\times10^{-3}$  &$\text{cm}~\text{s}^{-1}$ & SBL piston velocity of $\ch{CH4}$ &\citetalias{sauterey2020} \\
			$v_{\ch{CO}}$  & $4.8\times10^{-3}$  &$\text{cm}~\text{s}^{-1}$ & SBL piston velocity of $\ch{CO}$ &\citetalias{sauterey2020} \\
			$\alpha_{\ch{H2}}$  & $7.8\times10^{-4}$  &$\text{mol}~\text{L}^{-1}~\text{bar}^{-1} $& Henry coefficient for $\ch{H2}$ &\citetalias{sauterey2020} \\
			$\alpha_{\ch{CO2}}$ & Supp. Methods \ref{appendix:co2}  &$\text{mol}~\text{L}^{-1}~\text{bar}^{-1} $& Henry coefficient for $\ch{CO2}$ &\citetalias{sauterey2020} \\
			$\alpha_{\ch{CH4}}$ & $1.4\times10^{-3}$  &$\text{mol}~\text{L}^{-1}~\text{bar}^{-1} $& Henry coefficient for $\ch{CH4}$ &\citetalias{sauterey2020} \\
			$\alpha_{\ch{CO}}$  & $1 \times 10^{-3}$  &$\text{mol}~\text{L}^{-1}~\text{bar}^{-1} $& Henry coefficient for $\ch{CO}$ &\citetalias{sauterey2020} \\
			$C$ & $6.02\times 10^{20}$  &$\text{L}~\text{mol}^{-1}~\text{cm}^{-3}$ & Unit conversion constant &  \\
			$P_{tot}$ & 1  &bar & Total atmospheric pressure & assumed \\
			$S_p$     & $5.1\times 10^{18}$  &cm$^{2}$ & Planetary surface area & \\
			$M_{air}$ & 28.16 $\times 10^{-3}$  &$\text{kg}~\text{mol}^{-1}$ & Air molar mass &\citetalias{sauterey2020} \\
			$n_{tot}$ & $P_{tot}S_p\times 10^{-5}/(g M_{air})$  &mol & Total air mol &  \\
			$g$       & $9.80$   &m~s$^{-2}$ & Earth's gravitational acceleration &  \\
			$pK_w$ & 14  && Water ionic constant &  \\
			$M_o$ & $3\times 10^{21}$  &kg & Ocean mass & \\
			$M_d$ & $0.01 \times M_o$  &kg & Ocean mass & \\
			$J$ & $6.8\times 10^{16}$  &kg~yr$^{-1}$ & Ocean mixing rate & \citetalias{krissansen2017constraining}\\
			$k_{hyd}$ & $\exp{-\frac{10,570}{T+25.6}}$  &$\text{mol}^{-1}~\text{s}^{-1}$ & $\ch{CO}$ hydration rate coefficient &\citetalias{kharecha2005coupled} \\
			$k_2$ & $6.4\times 10^{-5}$  &$\text{s}^{-1}$ & Rate constant of \ch{HCOO- -> CO} &\citetalias{kharecha2005coupled} \\
			$k_3$ & $2.7 \times 10^{-6}$  &$\text{s}^{-1}$ & Rate constant of $\ch{HCOO- -> CH3COO-}$ &\citetalias{kharecha2005coupled} \\
			$T_{ME}$ & $15$  &$^{\circ}$C & surface temperature of the modern Earth & \citetalias{krissansen2017constraining}\\
            $f_{land}^{ME}$ & $0.29$  && modern Earth emerged land fraction & \citetalias{krissansen2017constraining}\\
			$K_{sol}$ & Supp. Methods \ref{appendix:solubilityproduct}  &$\text{mol}^{2}~\text{kg}^{-2}$ & Solubility product of carbonates &\citetalias{krissansen2017constraining}\\
			$k_{shelf}$ &  $1.03 \times 10^{13}$  &mol~yr$^{-1}$ & above shelf reference precipitation rate &\citetalias{krissansen2017constraining}\\
			$k_{pelagic}$ &  $1.40 \times 10^{12}$  &mol~yr$^{-1}$ & pelagic reference precipitation rate &\citetalias{krissansen2017constraining}\\
			$k_{p}$ &  $5.12 \times 10^{11}$  &mol~yr$^{-1}$ & deep ocean precipitation kinetic constant &\citetalias{krissansen2017constraining}\\
			$pH^{ME}$ & 8.2  && Modern pH of Earth's ocean & \\
			$Serp^{ME}$ & $10^{11}$  &mol~yr$^{-1}$ & Modern \ch{CH4} production by serpentinization &\citetalias{sauterey2020}\\
			$Volc^{ME}$ & $5.35\times 10^{12}$  &mol~yr$^{-1}$ & Modern volcanic outgassing of $\ch{H2}$ &\citetalias{sauterey2020}\\
            $k_{diss}$ & $F_{diss}^{ME} S_p \times e^{\frac{E_{bas}}{RT_p^{ME}}}$ & mol~yr$^{-1}$ & Basalt dissolution rate constant & \citetalias{krissansen2017constraining}$^{\dagger}$
\enddata
\end{deluxetable}}
\newpage
\section{Chemical equilibria}
\label{chemeq_appendix}
\subsection{Solubility of \ch{CO2}}
\label{appendix:co2}
From \citet{krissansen2017constraining}, we use the following parameterization for the Henry's law constant of $\ch{CO2}$ solubility as a function of temperature.
\begin{equation}
    \label{appendix:eq:co2henry}
    \begin{split}
    \alpha_{\ch{CO2}} = 10^{-5} \exp \{ & (\frac{9345.17}{T}-167.8108+23.3585 \log_e T \\
                                        &+(0.023517-2.3656\times10^{-4}T+4.7036\times 10^{-7}T^2)35.0) \}
    \end{split}
\end{equation}
$\text{mol}~\text{L}^{-1}~\text{bar}^{-1}$.

\subsection{Solubility product of carbonates in the ocean}
\label{appendix:solubilityproduct}
The solubility product of the ocean is obtained from \citet{krissansen2017constraining}, assuming a salinity of 35 per thousand :
\begin{equation}
\label{appendix:eq:solprod}
\begin{split}
\log_{10}K_{sol} = & -171.9065-0.077993T+2839.319/T+71.595\log_{10}(T) \\
                   & + (-0.77712+0.0028426T+178.34/T) 35^{0.5}\\
                   & - 0.0711 \times 35.0+0.0041249 \times 35^{1.5}
\end{split}
\end{equation}

\subsection{Carbonate equilibrium constants}
\label{appendix:carbeq}
The dissociation constants of $\ch{CO2}$ and $\ch{HCO3}^-$ are given as functions of temperatures (with a salinity of 35 per thousand:
\begin{equation}
    \log_{10}K_1(T) = -17.788+0.073104 \times T + 0.0051087 \times 35-1.1463e-4 T^2
\end{equation}
and
\begin{equation}
    \log_{10}K_2(T) = -20.919+0.064209 \times T + 0.011887 \times 35-8.7313e-5 T^2
\end{equation}

\section{Multiple linear regression  for atmospheric \ch{CO2}}
\label{sec:app:mreg}
We performed linear regression between $\log_{10}$ p\ch{CO2} and the input variables that cause variability in our system (luminosity, $S/\ssun$, emerged land fraction $f_{land}$, age, initial composition of the simulated atmosphere, initial pH See Figure~\ref{fig:decomposition}).

In our set of "mobile-lid" interior convection regime simulations, luminosity is the most variance explaining log-linear factor ($r^2 = 0.72$), then, emerged land fraction and planetary age (proxy to outgassing rate) explain most of the remaining variance ($r^2=0.14$ and $r^2=0.07$ respectively; Figure~\ref{fig:decomposition}A).
Not all the variance in the synthetic sample is explained by the variance-introducing variables (the multilinear $r^2$-value is found to be about 0.97 for the mobile-lid convection regime simulations).
This is due to some variance being introduced by discarded simulations, and to the relation between the variables and p\ch{CO2} not being exactly log-linear.

Relative luminosity being the most determining factor of atmospheric \ch{CO2} highlights that the carbonate-silicate cycle is at the core of the possibility for temperate conditions to exist throughout the habitable zone.
However, the importance of age and emerged land fraction seems to often be omitted in astronomical definitions of the HZ \citep[despite being actively discussed in the field of geophysics][]{Lenardic2016}.
This importance comes naturally from the fact that the geological carbon cycle model that we use \citep[taken from][]{krissansen2017constraining} is dependent on planetary age through the value of volcanic outgassing used in equations \ref{eq:fcarb}, \ref{eq:fsil}, and \ref{eq:fdiss}, and on the value of the emerged land fraction (assumed to be composed of continental material in the case of mobile-lid convection simulations) through equations \ref{eq:fcarb}, \ref{eq:fsil}.

In stagnant-lid convection simulations, luminosity is \emph{not} the variable with the highest correlation coefficient for $\log_{10}$p\ch{CO2}, instead, p\ch{CO2}$^0$ the initial \ch{CO2} pressure scores the highest correlation coefficient (Figure~\ref{fig:decomposition}B).
In SL simulations, emerged land does not contribute to weathering (Methods), hence SL simulations not showing a linear component in the $f_{land}$-p\ch{CO2} space (Figure~\ref{fig:decomposition}).
As luminosity negatively correlate with equilibrium p\ch{CO2} (Figure~\ref{fig:decomposition}D-F), p\ch{CO2}$^0$ \emph{positively} correlates with equilibrium p\ch{CO2} for stagnant-lid regime simulations (Figure~\ref{fig:decomposition}E).

The log-log correlation between p\ch{CO2} and p\ch{CO2}$^0$ in the dataset composed of simulations in the stagnant-lid interior convection regime is absent in the subset of these simulations of age below 2~Gyrs (Figure~\ref{fig:decomposition}C).

Others have suggested that age-based target selection could reduce required sample sizes for \textit{e.g.} biosignature assessment \citep{bixel2020testing}.
Our simulations further support this claim, as we find that age and the variables related to it is the secondmost correlating factor to atmospheric \ch{CO2} for mobile-lid regime simulations, and as the distribution of equilibrium p\ch{CO2} for stagnant-lid regime simulations is qualitatively different for 'old' and 'young' planets.
Hence, including age in a trend inference framework similar to the one we propose in the Main Text could result in fewer required planet in a sample to infer the dominant convection regime.
Because 'old' stagnant-lid regime simulations have a p\ch{CO2}--$S/\ssun$ distribution qualitatively different from that of mobile-lid regime simulations (whereas 'young' mobile-lid regime simulations have a similar distribution), an optimal strategy aimed at inferring the most common interior convection regime in Earth-like exoplanets would probably preferentially target 'old' ($>2$~Gyrs) targets.




\bibliography{main}{}

\begin{thebibliography}{}
\expandafter\ifx\csname natexlab\endcsname\relax\def\natexlab#1{#1}\fi
\providecommand{\url}[1]{\href{#1}{#1}}
\providecommand{\dodoi}[1]{doi:~\href{http://doi.org/#1}{\nolinkurl{#1}}}
\providecommand{\doeprint}[1]{\href{http://ascl.net/#1}{\nolinkurl{http://ascl.net/#1}}}
\providecommand{\doarXiv}[1]{\href{https://arxiv.org/abs/#1}{\nolinkurl{https://arxiv.org/abs/#1}}}

\bibitem[{Alt \& Teagle(1999)}]{alt1999uptake}
Alt, J.~C., \& Teagle, D.~A. 1999, Geochimica et Cosmochimica Acta, 63, 1527,
  \dodoi{10.1016/s0016-7037(99)00123-4}

\bibitem[{{Apai} {et~al.}(2019){Apai}, {Milster}, {Kim}, {Bixel}, {Schneider},
  {Liang}, \& {Arenberg}}]{apai2019nautilus}
{Apai}, D., {Milster}, T.~D., {Kim}, D.~W., {et~al.} 2019, \aj, 158, 83,
  \dodoi{10.3847/1538-3881/ab2631}

\bibitem[{{Apai} {et~al.}(2018){Apai}, {Ciesla}, {Mulders}, {Pascucci},
  {Barry}, {Pontoppidan}, {Bergin}, {Bixel}, {Brittain}, {Domagal-Goldman},
  {Hasegawa}, {Jang-Condell}, {Malhotra}, {Meyer}, {Youdin}, {Teske}, \&
  {Turner}}]{Apai2018}
{Apai}, D., {Ciesla}, F., {Mulders}, G.~D., {et~al.} 2018, arXiv e-prints,
  arXiv:1803.08682, \dodoi{10.48550/arXiv.1803.08682}

\bibitem[{Arney {et~al.}(2016)Arney, Domagal-Goldman, Meadows, Wolf,
  Schwieterman, Charnay, Claire, H{\'{e}}brard, \& Trainer}]{arney2016pale}
Arney, G., Domagal-Goldman, S.~D., Meadows, V.~S., {et~al.} 2016, Astrobiology,
  16, 873, \dodoi{10.1089/ast.2015.1422}

\bibitem[{Barber {et~al.}(1996)Barber, Dobkin, \&
  Huhdanpaa}]{barber1996quickhull}
Barber, C.~B., Dobkin, D.~P., \& Huhdanpaa, H. 1996, {ACM} Transactions on
  Mathematical Software, 22, 469, \dodoi{10.1145/235815.235821}

\bibitem[{Bar‐Nun \& Chang(1983)}]{BarNun1983}
Bar‐Nun, A., \& Chang, S. 1983, Journal of Geophysical Research: Oceans, 88,
  6662–6672, \dodoi{10.1029/jc088ic11p06662}

\bibitem[{Bean {et~al.}(2017)Bean, Abbot, \& Kempton}]{Bean2017}
Bean, J.~L., Abbot, D.~S., \& Kempton, E. M.-R. 2017, The Astrophysical
  Journal, 841, L24, \dodoi{10.3847/2041-8213/aa738a}

\bibitem[{Bixel \& Apai(2020)}]{bixel2020testing}
Bixel, A., \& Apai, D. 2020, The Astrophysical Journal, 896, 131,
  \dodoi{10.3847/1538-4357/ab8fad}

\bibitem[{Bixel \& Apai(2021)}]{bixel2021bioverse}
---. 2021, The Astronomical Journal, 161, 228, \dodoi{10.3847/1538-3881/abe042}

\bibitem[{Catling \& Kasting(2017)}]{catling2017atmospheric}
Catling, D.~C., \& Kasting, J.~F. 2017, Atmospheric Evolution on Inhabited and
  Lifeless Worlds (Cambridge University Press), \dodoi{10.1017/9781139020558}

\bibitem[{Catling {et~al.}(2018{\natexlab{a}})Catling, Krissansen-Totton,
  Kiang, Crisp, Robinson, DasSarma, Rushby, Genio, Bains, \&
  Domagal-Goldman}]{catling2018exoplanet}
Catling, D.~C., Krissansen-Totton, J., Kiang, N.~Y., {et~al.}
  2018{\natexlab{a}}, Astrobiology, 18, 709, \dodoi{10.1089/ast.2017.1737}

\bibitem[{Catling {et~al.}(2018{\natexlab{b}})Catling, Krissansen-Totton,
  Kiang, Crisp, Robinson, DasSarma, Rushby, Genio, Bains, \&
  Domagal-Goldman}]{catlingdavid2018exoplanet}
---. 2018{\natexlab{b}}, Astrobiology, 18, 709, \dodoi{10.1089/ast.2017.1737}

\bibitem[{Charnay {et~al.}(2013)Charnay, Forget, Wordsworth, Leconte, Millour,
  Codron, \& Spiga}]{charnay2013exploring}
Charnay, B., Forget, F., Wordsworth, R., {et~al.} 2013, Journal of Geophysical
  Research: Atmospheres, 118, 10,414,
  \dodoi{https://doi.org/10.1002/jgrd.50808}

\bibitem[{Cockell {et~al.}(2016)Cockell, Bush, Bryce, Direito, Fox-Powell,
  Harrison, Lammer, Landenmark, Martin-Torres, Nicholson, Noack,
  O{\textquotesingle}Malley-James, Payler, Rushby, Samuels, Schwendner,
  Wadsworth, \& Zorzano}]{cockell2016habitability}
Cockell, C., Bush, T., Bryce, C., {et~al.} 2016, Astrobiology, 16, 89,
  \dodoi{10.1089/ast.2015.1295}

\bibitem[{Dorn {et~al.}(2018)Dorn, Noack, \& Rozel}]{dorn2018outgassing}
Dorn, C., Noack, L., \& Rozel, A.~B. 2018, Astronomy {\&} Astrophysics, 614,
  A18, \dodoi{10.1051/0004-6361/201731513}

\bibitem[{Fauchez {et~al.}(2021)Fauchez, Turbet, Sergeev, Mayne, Spiga, Sohl,
  Saxena, Deitrick, Gilli, Domagal-Goldman, Forget, Consentino, Barnes,
  Haqq-Misra, Way, Wolf, Olson, Crouse, Janin, Bolmont, Leconte, Chaverot,
  Jaziri, Tsigaridis, Yang, Pidhorodetska, Kopparapu, Chen, Boutle, Lefevre,
  Charnay, Burnett, Cabra, \& Bouldin}]{Fauchez2021}
Fauchez, T.~J., Turbet, M., Sergeev, D.~E., {et~al.} 2021, The Planetary
  Science Journal, 2, 106, \dodoi{10.3847/psj/abf4df}

\bibitem[{Foley(2019)}]{foley2019habitability}
Foley, B.~J. 2019, The Astrophysical Journal, 875, 72,
  \dodoi{10.3847/1538-4357/ab0f31}

\bibitem[{Foley \& Smye(2018)}]{foley2018carbon}
Foley, B.~J., \& Smye, A.~J. 2018, Astrobiology, 18, 873,
  \dodoi{10.1089/ast.2017.1695}

\bibitem[{{Gaudi} {et~al.}(2020){Gaudi}, {Seager}, {Mennesson}, {Kiessling},
  {Warfield}, {Cahoy}, {Clarke}, {Domagal-Goldman}, {Feinberg}, {Guyon},
  {Kasdin}, {Mawet}, {Plavchan}, {Robinson}, {Rogers}, {Scowen}, {Somerville},
  {Stapelfeldt}, {Stark}, {Stern}, {Turnbull}, {Amini}, {Kuan}, {Martin},
  {Morgan}, {Redding}, {Stahl}, {Webb}, {Alvarez-Salazar}, {Arnold}, {Arya},
  {Balasubramanian}, {Baysinger}, {Bell}, {Below}, {Benson}, {Blais}, {Booth},
  {Bourgeois}, {Bradford}, {Brewer}, {Brooks}, {Cady}, {Caldwell}, {Calvet},
  {Carr}, {Chan}, {Cormarkovic}, {Coste}, {Cox}, {Danner}, {Davis}, {Dewell},
  {Dorsett}, {Dunn}, {East}, {Effinger}, {Eng}, {Freebury}, {Garcia}, {Gaskin},
  {Greene}, {Hennessy}, {Hilgemann}, {Hood}, {Holota}, {Howe}, {Huang}, {Hull},
  {Hunt}, {Hurd}, {Johnson}, {Kissil}, {Knight}, {Kolenz}, {Kraus}, {Krist},
  {Li}, {Lisman}, {Mandic}, {Mann}, {Marchen}, {Marrese-Reading}, {McCready},
  {McGown}, {Missun}, {Miyaguchi}, {Moore}, {Nemati}, {Nikzad}, {Nissen},
  {Novicki}, {Perrine}, {Pineda}, {Polanco}, {Putnam}, {Qureshi}, {Richards},
  {Eldorado Riggs}, {Rodgers}, {Rud}, {Saini}, {Scalisi}, {Scharf}, {Schulz},
  {Serabyn}, {Sigrist}, {Sikkia}, {Singleton}, {Shaklan}, {Smith}, {Southerd},
  {Stahl}, {Steeves}, {Sturges}, {Sullivan}, {Tang}, {Taras}, {Tesch},
  {Therrell}, {Tseng}, {Valente}, {Van Buren}, {Villalvazo}, {Warwick}, {Webb},
  {Westerhoff}, {Wofford}, {Wu}, {Woo}, {Wood}, {Ziemer}, {Arney}, {Anderson},
  {Ma{\'\i}z-Apell{\'a}niz}, {Bartlett}, {Belikov}, {Bendek}, {Cenko},
  {Douglas}, {Dulz}, {Evans}, {Faramaz}, {Feng}, {Ferguson}, {Follette},
  {Ford}, {Garc{\'\i}a}, {Geha}, {Gelino}, {G{\"o}tberg}, {Hildebrandt}, {Hu},
  {Jahnke}, {Kennedy}, {Kreidberg}, {Isella}, {Lopez}, {Marchis}, {Macri},
  {Marley}, {Matzko}, {Mazoyer}, {McCandliss}, {Meshkat}, {Mordasini},
  {Morris}, {Nielsen}, {Newman}, {Petigura}, {Postman}, {Reines}, {Roberge},
  {Roederer}, {Ruane}, {Schwieterman}, {Sirbu}, {Spalding}, {Teplitz},
  {Tumlinson}, {Turner}, {Werk}, {Wofford}, {Wyatt}, {Young}, \&
  {Zellem}}]{gaudi2020habitable}
{Gaudi}, B.~S., {Seager}, S., {Mennesson}, B., {et~al.} 2020, arXiv e-prints,
  arXiv:2001.06683, \dodoi{10.48550/arXiv.2001.06683}

\bibitem[{Gillis \& Coogan(2011)}]{gillis2011secular}
Gillis, K., \& Coogan, L. 2011, Earth and Planetary Science Letters, 302, 385,
  \dodoi{10.1016/j.epsl.2010.12.030}

\bibitem[{Haqq-Misra {et~al.}(2008)Haqq-Misra, Domagal-Goldman, Kasting, \&
  Kasting}]{haqq2008revised}
Haqq-Misra, J.~D., Domagal-Goldman, S.~D., Kasting, P.~J., \& Kasting, J.~F.
  2008, Astrobiology, 8, 1127, \dodoi{10.1089/ast.2007.0197}

\bibitem[{Hart(1979)}]{Hart1979}
Hart, M.~H. 1979, Icarus, 37, 351–357, \dodoi{10.1016/0019-1035(79)90141-6}

\bibitem[{Hessler {et~al.}(2004)Hessler, Lowe, Jones, \& Bird}]{Hessler2004}
Hessler, A.~M., Lowe, D.~R., Jones, R.~L., \& Bird, D.~K. 2004, Nature, 428,
  736–738, \dodoi{10.1038/nature02471}

\bibitem[{Huang(1959)}]{huang1959occurrence}
Huang, S.-S. 1959, American scientist, 47, 397,
  \dodoi{http://www.jstor.org/stable/27827376}

\bibitem[{Jeffreys(1967)}]{jeffreys1967theory}
Jeffreys, H. 1967, Theory of probability., 3rd edn. (Oxford: Clarendon Press)

\bibitem[{Kanzaki \& Murakami(2015)}]{Kanzaki2015}
Kanzaki, Y., \& Murakami, T. 2015, Geochimica et Cosmochimica Acta, 159,
  190–219, \dodoi{10.1016/j.gca.2015.03.011}

\bibitem[{Kass \& Raftery(1995)}]{kass1995bayes}
Kass, R.~E., \& Raftery, A.~E. 1995, Journal of the American Statistical
  Association, 90, 773, \dodoi{10.1080/01621459.1995.10476572}

\bibitem[{Kasting(1993)}]{Kasting1993Earth}
Kasting, J.~F. 1993, Science, 259, 920–926, \dodoi{10.1126/science.11536547}

\bibitem[{Kasting {et~al.}(1993)Kasting, Whitmire, \&
  Reynolds}]{kasting1993habitable}
Kasting, J.~F., Whitmire, D.~P., \& Reynolds, R.~T. 1993, Icarus, 101, 108,
  \dodoi{10.1006/icar.1993.1010}

\bibitem[{Kharecha {et~al.}(2005)Kharecha, Kasting, \&
  Siefert}]{kharecha2005coupled}
Kharecha, P., Kasting, J., \& Siefert, J. 2005, Geobiology, 3, 53,
  \dodoi{10.1111/j.1472-4669.2005.00049.x}

\bibitem[{Kopparapu {et~al.}(2013)Kopparapu, Ramirez, Kasting, Eymet, Robinson,
  Mahadevan, Terrien, Domagal-Goldman, Meadows, \&
  Deshpande}]{kopparapu2013habitable}
Kopparapu, R.~K., Ramirez, R., Kasting, J.~F., {et~al.} 2013, The Astrophysical
  Journal, 765, 131, \dodoi{10.1088/0004-637x/765/2/131}

\bibitem[{Krissansen-Totton {et~al.}(2018)Krissansen-Totton, Arney, \&
  Catling}]{krissansen2018constraining}
Krissansen-Totton, J., Arney, G.~N., \& Catling, D.~C. 2018, Proceedings of the
  National Academy of Sciences, 115, 4105, \dodoi{10.1073/pnas.1721296115}

\bibitem[{Krissansen-Totton \& Catling(2017)}]{krissansen2017constraining}
Krissansen-Totton, J., \& Catling, D.~C. 2017, Nature Communications, 8,
  \dodoi{10.1038/ncomms15423}

\bibitem[{Krissansen-Totton {et~al.}(2022)Krissansen-Totton, Thompson,
  Galloway, \& Fortney}]{krissansen2022understanding}
Krissansen-Totton, J., Thompson, M., Galloway, M.~L., \& Fortney, J.~J. 2022,
  Nature Astronomy, 6, 189, \dodoi{10.1038/s41550-021-01579-7}

\bibitem[{Lehmer {et~al.}(2020)Lehmer, Catling, \&
  Krissansen-Totton}]{lehmer2020carbonate}
Lehmer, O.~R., Catling, D.~C., \& Krissansen-Totton, J. 2020, Nature
  Communications, 11, \dodoi{10.1038/s41467-020-19896-2}

\bibitem[{Lenardic {et~al.}(2016)Lenardic, Jellinek, Foley, O’Neill, \&
  Moore}]{Lenardic2016}
Lenardic, A., Jellinek, A.~M., Foley, B., O’Neill, C., \& Moore, W.~B. 2016,
  Journal of Geophysical Research: Planets, 121, 1831–1864,
  \dodoi{10.1002/2016je005089}

\bibitem[{Mamajek \& Stapelfeldt(2023)}]{mamajek2023}
Mamajek, E., \& Stapelfeldt, K. 2023, Mission Star List for the {H}abitable
  {W}orlds {O}bservatory,  {NASA} Jet Propulsion Laboratory

\bibitem[{Manabe \& Wetherald(1967)}]{manabe1967thermal}
Manabe, S., \& Wetherald, R.~T. 1967, Journal of the Atmospheric Sciences, 24,
  241, \dodoi{10.1175/1520-0469(1967)024<0241:teotaw>2.0.co;2}

\bibitem[{Mazevet {et~al.}(2023)Mazevet, Affholder, Sauterey, Bixel, Apai, \&
  Ferriere}]{mazevet2023prospects}
Mazevet, S., Affholder, A., Sauterey, B., {et~al.} 2023, Comptes Rendus.
  Physique, \dodoi{10.5802/crphys.154}

\bibitem[{Meadows {et~al.}(2018)Meadows, Reinhard, Arney, Parenteau,
  Schwieterman, Domagal-Goldman, Lincowski, Stapelfeldt, Rauer, DasSarma,
  Hegde, Narita, Deitrick, Lustig-Yaeger, Lyons, Siegler, \&
  Grenfell}]{meadows2018exoplanet}
Meadows, V.~S., Reinhard, C.~T., Arney, G.~N., {et~al.} 2018, Astrobiology, 18,
  630, \dodoi{10.1089/ast.2017.1727}

\bibitem[{{Moresi} \& {Solomatov}(1998)}]{moresi1998}
{Moresi}, L., \& {Solomatov}, V. 1998, Geophysical Journal International, 133,
  669, \dodoi{10.1046/j.1365-246X.1998.00521.x}

\bibitem[{{National Academies of Sciences, Engineering, and
  Medicine}(2021)}]{NAP26141}
{National Academies of Sciences, Engineering, and Medicine}. 2021, Pathways to
  Discovery in Astronomy and Astrophysics for the 2020s (Washington, DC: The
  National Academies Press), \dodoi{10.17226/26141}

\bibitem[{{Rauer} {et~al.}(2024){Rauer}, {Aerts}, {Cabrera}, {Deleuil},
  {Erikson}, {Gizon}, {Goupil}, {Heras}, {Lorenzo-Alvarez}, {Marliani},
  {Martin-Garcia}, {Mas-Hesse}, {O'Rourke}, {Osborn}, {Pagano}, {Piotto},
  {Pollacco}, {Ragazzoni}, {Ramsay}, {Udry}, {Appourchaux}, {Benz},
  {Brandeker}, {G{\"u}del}, {Janot-Pacheco}, {Kabath}, {Kjeldsen}, {Min},
  {Santos}, {Smith}, {Suarez}, {Werner}, {Aboudan}, {Abreu}, {Acu{\~n}a},
  {Adams}, {Adibekyan}, {Affer}, {Agneray}, {Agnor}, {Aguirre B{\o}rsen-Koch},
  {Ahmed}, {Aigrain}, {Al-Bahlawan}, {Alcacera Gil}, {Alei}, {Alencar},
  {Alexander}, {Alfonso-Garz{\'o}n}, {Alibert}, {Allende Prieto}, {Almeida},
  {Alonso Sobrino}, {Altavilla}, {Althaus}, {Alonso Alvarez Trujillo},
  {Amarsi}, {Ammler-von Eiff}, {Am{\^o}res}, {Andrade}, {Antoniadis-Karnavas},
  {Ant{\'o}nio}, {Aparicio del Moral}, {Appolloni}, {Arena}, {Armstrong},
  {Aroca Aliaga}, {Asplund}, {Audenaert}, {Auricchio}, {Avelino}, {Baeke},
  {Bailli{\'e}}, {Balado}, {Balestra}, {Ball}, {Ballans}, {Ballot}, {Barban},
  {Barbary}, {Barbieri}, {Barcel{\'o} Forteza}, {Barker}, {Barklem}, {Barnes},
  {Barrado Navascues}, {Barragan}, {Baruteau}, {Basu}, {Baudin}, {Baumeister},
  {Bayliss}, {Bazot}, {Beck}, {Bedding}, {Belkacem}, {Bellinger}, {Benatti},
  {Benomar}, {B{\'e}rard}, {Bergemann}, {Bergomi}, {Bernardo}, {Biazzo},
  {Bignamini}, {Bigot}, {Billot}, {Binet}, {Biondi}, {Biondi}, {Birch},
  {Bitsch}, {Bluhm Ceballos}, {B{\'o}di}, {Bogn{\'a}r}, {Boisse}, {Bolmont},
  {Bonanno}, {Bonavita}, {Bonfanti}, {Bonfils}, {Bonito}, {Bonomo},
  {B{\"o}rner}, {Boro Saikia}, {Borreguero Mart{\'\i}n}, {Borsa}, {Borsato},
  {Bossini}, {Bouchy}, {Bou{\'e}}, {Boufleur}, {Boumier}, {Bourrier}, {Bowman},
  {Bozzo}, {Bradley}, {Bray}, {Bressan}, {Breton}, {Brienza}, {Brito}, {Brogi},
  {Brown}, {Brown}, {Brun}, {Bruno}, {Bruns}, {Buchhave}, {Bugnet}, {Buldgen},
  {Burgess}, {Busatta}, {Busso}, {Buzasi}, {Caballero}, {Cabral}, {Calderone},
  {Cameron}, {Cameron}, {Campante}, {Canto Martins}, {Cara}, {Carone},
  {Carrasco}, {Casagrande}, {Casewell}, {Cassisi}, {Castellani}, {Castro},
  {Catala}, {Catal{\'a}n Fern{\'a}ndez}, {Catelan}, {Cegla}, {Cerruti},
  {Cessa}, {Chadid}, {Chaplin}, {Charpinet}, {Chiappini}, {Chiarucci},
  {Chiavassa}, {Chinellato}, {Chirulli}, {Christensen-Dalsgaard}, {Church},
  {Claret}, {Clarke}, {Claudi}, {Clermont}, {Coelho}, {Coelho}, {Cogato},
  {Colom{\'e}}, {Condamin}, {Conseil}, {Corbard}, {Correia}, {Corsaro},
  {Cosentino}, {Costes}, {Cottinelli}, {Covone}, {Creevey}, {Crida},
  {Csizmadia}, {Cunha}, {Curry}, {da Costa}, {da Silva}, {Dalal}, {Damasso},
  {Damiani}, {Damiani}, {Liduina das Chagas}, {Davies}, {Davies}, {Davies},
  {Davison}, {de Almeida}, {de Angeli}, {Cabral de Barros}, {de Castro
  Le{\~a}o}, {Brito de Freitas}, {de Freitas}, {De Martino}, {Renan de
  Medeiros}, {de Paula}, {de Plaa}, {De Ridder}, {Deal}, {Decin}, {Deeg},
  {Degl'Innocenti}, {Deheuvels}, {del Burgo}, {Del Sordo}, {Delgado-Mena},
  {Demangeon}, {Denk}, {Derekas}, {Desidera}, {Dexet}, {Di Criscienzo}, {Di
  Giorgio}, {Di Mauro}, {Diaz Rial}, {D{\'\i}az-Garc{\'\i}a}, {Dima},
  {Dinuzzi}, {Dionatos}, {Distefano}, {do Nascimento}, {Domingo}, {D'Orazi},
  {Dorn}, {Doyle}, {Duarte}, {Ducellier}, {Dumaye}, {Dumusque}, {Dupret},
  {Eggenberger}, {Ehrenreich}, {Eigm{\"u}ller}, {Eising}, {Emilio}, {Eriksson},
  {Ermocida}, {Isidoro Escate Giribaldi}, {Eschen}, {Estrela}, {Evans},
  {Fabbian}, {Fabrizio}, {Faria}, {Farina}, {Farinato}, {Feliz}, {Feltzing},
  {Fenouillet}, {Ferrari}, {Ferraz-Mello}, {Fialho}, {Fienga}, {Figueira},
  {Fiori}, {Flaccomio}, {Focardi}, {Foley}, {Fontignie}, {Ford}, {Fornazier},
  {Forveille}, {Fossati}, {de Marca Franca}, {da Silva}, {Frasca}, {Fridlund},
  {Furlan}, {Gabler}, {Gaido}, {Gallagher}, {Galli}, {Garcia}, {Garc{\'\i}a
  Hern{\'a}ndez}, {Garcia Munoz}, {Garc{\'\i}a-V{\'a}zquez}, {Garrido Haba},
  {Gaulme}, {Gauthier}, {Gehan}, {Gent}, {Georgieva}, {Ghigo}, {Giana}, {Gill},
  {Girardi}, {Giuliatti Winter}, {Giusi}, {Gomes da Silva}, {G{\'o}mez Zazo},
  {Gomez-Lopez}, {Isai Gonz{\'a}lez Hern{\'a}ndez}, {Gonzalez Murillo},
  {Gorius}, {Gouel}, {Goulty}, {Granata}, {Grenfell}, {Grie{\ss}bach},
  {Grolleau}, {Grouffal}, {Grziwa}, {Guarcello}, {Gueguen}, {Guenther},
  {Guilhem}, {Guillerot}, {Guiot}, {Guterman}, {Guti{\'e}rrez},
  {Guti{\'e}rrez-Canales}, {Hagelberg}, {Haldemann}, {Hall}, {Handberg},
  {Harrison}, {Harrison}, {Hasiba}, {Haswell}, {Hatalova}, {Hatzes}, {Haywood},
  {H{\'e}brard}, {Heckes}, {Heiter}, {Hekker}, {Heller}, {Helling},
  {Helminiak}, {Hemsley}, {Heng}, {Hermans}, {Hermes}, {Hidalgo Torres},
  {Hinkel}, {Hobbs}, {Hodgkin}, {Hofmann}, {Hojjatpanah}, {Houdek}, {Huber},
  {Huesler}, {Hui-Bon-Hoa}, {Huygen}, {Huynh}, {Iro}, {Irwin}, {Irwin},
  {Izidoro}, {Jacquinod}, {Emborg Jannsen}, {Janson}, {Jeszenszky}, {Jiang},
  {Jos{\'e} Jimenez Mancebo}, {Jofre}, {Johansen}, {Johnston}, {Jones},
  {Kallinger}, {K{\'a}lm{\'a}n}, {Kanitz}, {Karjalainen}, {Karjalainen},
  {Karoff}, {Kawaler}, {Kawata}, {Keereman}, {Keiderling}, {Kennedy},
  {Kenworthy}, {Kerschbaum}, {Kidger}, {Kiefer}, {Kintziger}, {Kislyakova},
  {Kiss}, {Klagyivik}, {Klahr}, {Klevas}, {Kochukhov}, {K{\"o}hler}, {Kolb},
  {Koncz}, {Korth}, {Kostogryz}, {Kov{\'a}cs}, {Kov{\'a}cs}, {Kozhura},
  {Krivova}, {Ku{\v{c}}inskas}, {Kuhlemann}, {Kupka}, {Laauwen}, {Labiano},
  {Lagarde}, {Laget}, {Laky}, {Lam}, {Lambrechts}, {Lammer}, {Lanza},
  {Lanzafame}, {Lares Martiz}, {Laskar}, {Latter}, {Lavanant}, {Lawrenson},
  {Lazzoni}, {Lebre}, {Lebreton}, {Lecavelier des Etangs}, {Leinhardt},
  {Leleu}, {Lendl}, {Leto}, {Levillain}, {Libert}, {Lichtenberg}, {Ligi},
  {Lignieres}, {Lillo-Box}, {Linsky}, {Scige Liu}, {Loidolt}, {Longval},
  {Lopes}, {Lorenzani}, {Ludwig}, {Lund}, {Sloth Lundkvist}, {Luri},
  {Maceroni}, {Madden}, {Madhusudhan}, {Maggio}, {Magliano}, {Magrin}, {Mahy},
  {Maibaum}, {Malac-Allain}, {Malapert}, {Malavolta}, {Maldonado}, {Mamonova},
  {Manchon}, {Mann}, {Mantovan}, {Marafatto}, {Marconi}, {Mardling}, {Marigo},
  {Marinoni}, {Marques}, {Marques}, {Marrese}, {Marshall}, {Mart{\'\i}nez
  Perales}, {Mary}, {Marzari}, {Masana}, {Mascher}, {Mathis}, {Mathur},
  {Mattiuci Figueiredo}, {Maxted}, {Mazeh}, {Mazevet}, {Mazzei}, {McCormac},
  {McMillan}, {Menou}, {Merle}, {Meru}, {Mesa}, {Messina}, {M{\'e}sz{\'a}ros},
  {Meunier}, {Meunier}, {Micela}, {Michaelis}, {Michel}, {Michielsen},
  {Michtchenko}, {Miglio}, {Miguel}, {Milligan}, {Mirouh}, {Mitchel}, {Moedas},
  {Molendini}, {Moln{\'a}r}, {Mombarg}, {Montalban}, {Montalto}, {Monteiro},
  {Morales}, {Morales-Calderon}, {Morbidelli}, {Mordasini}, {Moreau}, {Morel},
  {Morello}, {Morin}, {Mortier}, {Mosser}, {Mourard}, {Mousis}, {Moutou},
  {Mowlavi}, {Moya}, {Muehlmann}, {Muirhead}, {Munari}, {Musella}, {Mustill},
  {Nardetto}, {Nardiello}, {Narita}, {Nascimbeni}, {Nash}, {Neiner}, {Nelson},
  {Nettelmann}, {Nicolini}, {Nielsen}, {Niemi}, {Noack}, {Noels-Grotsch},
  {Noll}, {Norazman}, {Norton}, {Nsamba}, {Ofir}, {Ogilvie}, {Olander},
  {Olivetto}, {Olofsson}, {Ong}, {Ortolani}, {Oshagh}, {Ottacher},
  {Ottensamer}, {Ouazzani}, {Paardekooper}, {Pace}, {Pajas}, {Palacios},
  {Palandri}, {Palle}, {Paproth}, {Parro}, {Parviainen}, {Granado},
  {Passegger}, {Pastor-Morales}, {P{\"a}tzold}, {Gade Pedersen}, {Pena
  Hidalgo}, {Pepe}, {Pereira}, {Persson}, {Pertenais}, {Peter}, {Petit},
  {Petit}, {Pezzuto}, {Pichierri}, {Pietrinferni}, {Pinheiro}, {Pinsonneault},
  {Plachy}, {Plasson}, {Plez}, {Poppenhaeger}, {Poretti}, {Portaluri},
  {Portell}, {Frederico Porto de Mello}, {Poyatos}, {Pozuelos}, {Prada Moroni},
  {Pricopi}, {Prisinzano}, {Quade}, {Quirrenbach160}, {Rabanal Reina6},
  {Rabello Soares}, {Raimondo}, {Rainer}, {Ram{\'o}n Rod{\'o}n},
  {Ram{\'o}n-Ballesta}, {Ramos Zapata}, {R{\"a}tz}, {Rauterberg}, {Redman},
  {Redmer}, {Reese}, {Regibo}, {Reiners}, {Reinhold}, {Renie}, {Ribas},
  {Ribeiro}, {Pereira Ricciardi}, {Rice}, {Richard}, {Riello}, {Rieutord},
  {Ripepi}, {Rixon}, {Rockstein}, {Rodr{\'\i}guez}, {Rodr{\'\i}guez D{\'\i}az},
  {Rodriguez Garcia}, {Rodriguez-Gomez}, {Roehlly}, {Roig}, {Rojas-Ayala},
  {Rolf}, {Lysgaard R{\o}rsted}, {Rosado}, {Rosotti}, {Roth}, {Roth},
  {Rousseau}, {Roxburgh}, {Roy}, {Royer}, {Ruane}, {Rufini Mastropasqua}, {Ruiz
  de Galarreta}, {Russi}, {Saar}, {Saillenfest}, {Salaris}, {Salmon}, {Saltas},
  {Samadi}, {Samadi}, {Samra}, {Sanches da Silva}, {Andr{\'e}s S{\'a}nchez
  Carrasco}, {Santerne}, {Santoli}, {Santos}, {Sanz Mesa}, {Sarro},
  {Scandariato}, {Sch{\"a}fer}, {Schlafly}, {Schmider}, {Schneider}, {Schou},
  {Schunker}, {J{\"o}rg Schwarzkopf}, {Serenelli}, {Seynaeve}, {Shan},
  {Shapiro}, {Shipman}, {Sicilia}, {Sierra Sanmartin}, {Sigot}, {Silliman},
  {Silvotti}, {Simon}, {Simoyama Napoli}, {Skarka}, {Smalley}, {Smiljanic},
  {Smit}, {Smith}, {Smith}, {Snellen}, {S{\'o}dor}, {Sohl}, {Solanki},
  {Sortino}, {Sousa}, {Southworth}, {Souto}, {Sozzetti}, {Stamatellos},
  {Stassun}, {Steller}, {Stello}, {Stelzer}, {Stiebeler}, {Stokholm},
  {Storelvmo}, {Strassmeier}, {Str{\o}m}, {Strugarek}, {Sulis}, {{\v{S}}vanda},
  {Szabados}, {Szab{\'o}}, {Szab{\'o}}, {Szuszkiewicz}, {Talens}, {Teti},
  {Theisen}, {Th{\'e}venin}, {Thoul}, {Tiphene}, {Titz-Weider}, {Tkachenko},
  {Tomecki}, {Tonfat}, {Tosi}, {Trampedach}, {Traven}, {Triaud}, {Tr{\o}nnes},
  {Tsantaki}, {Tschentscher}, {Turin}, {Tvaruzka}, {Ulmer}, {Ulmer-Moll},
  {Ulusoy}, {Umbriaco}, {Valencia}, {Valentini}, {Valio}, {Valverde Guijarro},
  {Van Eylen}, {Van Grootel}, {van Kempen}, {Van Reeth}, {Van Zelst},
  {Vandenbussche}, {Vasiliou}, {Vasilyev}, {Vaz de Mascarenhas}, {Vazan}, {Vela
  Nunez}, {Nunes Velloso}, {Ventura}, {Ventura}, {Venturini}, {Trallero},
  {Veras}, {Verdugo}, {Verma}, {Vibert}, {Vicanek Martinez}, {Vida}, {Vigan},
  {Villacorta}, {Villaver}, {Villaverde Aparicio}, {Viotto}, {Vorobyov},
  {Vorontsov}, {Wagner}, {Walloschek}, {Walton}, {Walton}, {Wang}, {Waters},
  {Watson}, {Wedemeyer}, {Weeks}, {Weingril}, {Weiss}, {Wendler}, {West},
  {Westerdorff}, {Westphal}, {Wheatley}, {White}, {Whittaker}, {Wickhusen},
  {Wilson}, {Windsor}, {Winter}, {Lykke Winther}, {Winton}, {Witteck},
  {Witzke}, {Woitke}, {Wolter}, {Wuchterl}, {Wyatt}, {Yang}, {Yu}, {Zanmar
  Sanchez}, {Rosa Zapatero Osorio}, {Zechmeister}, {Zhou}, {Ziemke}, \&
  {Zwintz}}]{PLATO2024}
{Rauer}, H., {Aerts}, C., {Cabrera}, J., {et~al.} 2024, arXiv e-prints,
  arXiv:2406.05447, \dodoi{10.48550/arXiv.2406.05447}

\bibitem[{Robinson(2018)}]{Robinson2018}
Robinson, T.~D. 2018, Characterizing Exoplanet Habitability (Springer
  International Publishing), 3137--3157, \dodoi{10.1007/978-3-319-55333-7_67}

\bibitem[{Rye {et~al.}(1995)Rye, Kuo, \& Holland}]{Rye1995}
Rye, R., Kuo, P.~H., \& Holland, H.~D. 1995, Nature, 378, 603–605,
  \dodoi{10.1038/378603a0}

\bibitem[{Sauterey {et~al.}(2020)Sauterey, Charnay, Affholder, Mazevet, \&
  Ferri{\`{e}}re}]{sauterey2020}
Sauterey, B., Charnay, B., Affholder, A., Mazevet, S., \& Ferri{\`{e}}re, R.
  2020, Nature Communications, 11, \dodoi{10.1038/s41467-020-16374-7}

\bibitem[{Sch\"{o}nbrodt \& Wagenmakers(2017)}]{schonbrodt2017bayes}
Sch\"{o}nbrodt, F.~D., \& Wagenmakers, E.-J. 2017, Psychonomic Bulletin {\&}
  Review, 25, 128, \dodoi{10.3758/s13423-017-1230-y}

\bibitem[{Sheldon(2006)}]{Sheldon2006}
Sheldon, N.~D. 2006, Precambrian Research, 147, 148–155,
  \dodoi{10.1016/j.precamres.2006.02.004}

\bibitem[{Sleep \& Zahnle(2001)}]{sleep2001}
Sleep, N.~H., \& Zahnle, K. 2001, Journal of Geophysical Research: Planets,
  106, 1373, \dodoi{https://doi.org/10.1029/2000JE001247}

\bibitem[{Solomatov \& Moresi(1996)}]{solomatov1996stagnant}
Solomatov, V.~S., \& Moresi, L.-N. 1996, Journal of Geophysical Research:
  Planets, 101, 4737, \dodoi{10.1029/95je03361}

\bibitem[{{The LUVOIR Team}(2019)}]{luvoir2019luvoir}
{The LUVOIR Team}. 2019, The LUVOIR Mission Concept Study Final Report,  arXiv,
  \dodoi{10.48550/ARXIV.1912.06219}

\bibitem[{Tosi {et~al.}(2017)Tosi, Godolt, Stracke, Ruedas, Grenfell,
  H\"{o}ning, Nikolaou, Plesa, Breuer, \& Spohn}]{Tosi2017}
Tosi, N., Godolt, M., Stracke, B., {et~al.} 2017, Astronomy \& Astrophysics,
  605, A71, \dodoi{10.1051/0004-6361/201730728}

\bibitem[{Villanueva {et~al.}(2018)Villanueva, Smith, Protopapa, Faggi, \&
  Mandell}]{Villanueva2018}
Villanueva, G., Smith, M., Protopapa, S., Faggi, S., \& Mandell, A. 2018,
  Journal of Quantitative Spectroscopy and Radiative Transfer, 217, 86,
  \dodoi{10.1016/j.jqsrt.2018.05.023}

\bibitem[{Virtanen {et~al.}(2020)Virtanen, Gommers, Oliphant, Haberland, Reddy,
  Cournapeau, Burovski, Peterson, Weckesser, Bright, van~der Walt, Brett,
  Wilson, Millman, Mayorov, Nelson, Jones, Kern, Larson, Carey, Polat, Feng,
  Moore, VanderPlas, Laxalde, Perktold, Cimrman, Henriksen, Quintero, Harris,
  Archibald, Ribeiro, Pedregosa, van Mulbregt, Vijaykumar, Bardelli, Rothberg,
  Hilboll, Kloeckner, Scopatz, Lee, Rokem, Woods, Fulton, Masson,
  H\"{a}ggstr\"{o}m, Fitzgerald, Nicholson, Hagen, Pasechnik, Olivetti, Martin,
  Wieser, Silva, Lenders, Wilhelm, Young, Price, Ingold, Allen, Lee, Audren,
  Probst, Dietrich, Silterra, Webber, Slavi{\v c}, Nothman, Buchner, Kulick,
  Sch\"{o}nberger, de~Miranda~Cardoso, Reimer, Harrington, Rodr{\'\i}guez,
  Nunez-Iglesias, Kuczynski, Tritz, Thoma, Newville, K\"{u}mmerer, Bolingbroke,
  Tartre, Pak, Smith, Nowaczyk, Shebanov, Pavlyk, Brodtkorb, Lee, McGibbon,
  Feldbauer, Lewis, Tygier, Sievert, Vigna, Peterson, More, Pudlik, Oshima,
  Pingel, Robitaille, Spura, Jones, Cera, Leslie, Zito, Krauss, Upadhyay,
  Halchenko, \& V{\'a}zquez-Baeza}]{virtanen2020scipy}
Virtanen, P., Gommers, R., Oliphant, T.~E., {et~al.} 2020, Nature Methods, 17,
  261, \dodoi{10.1038/s41592-019-0686-2}

\bibitem[{von Paris {et~al.}(2008)von Paris, Rauer, Grenfell, Patzer, Hedelt,
  Stracke, Trautmann, \& Schreier}]{von2008warming}
von Paris, P., Rauer, H., Grenfell, J.~L., {et~al.} 2008, Planetary and Space
  Science, 56, 1244, \dodoi{10.1016/j.pss.2008.04.008}

\bibitem[{Walker {et~al.}(1981)Walker, Hays, \& Kasting}]{walker1981negative}
Walker, J. C.~G., Hays, P.~B., \& Kasting, J.~F. 1981, Journal of Geophysical
  Research, 86, 9776, \dodoi{10.1029/jc086ic10p09776}

\bibitem[{Windley {et~al.}(2021)Windley, Kusky, \& Polat}]{Windley2021}
Windley, B.~F., Kusky, T., \& Polat, A. 2021, Precambrian Research, 352,
  105980, \dodoi{10.1016/j.precamres.2020.105980}

\bibitem[{Wolf \& Toon(2013)}]{wolf2013hospitable}
Wolf, E., \& Toon, O. 2013, Astrobiology, 13, 656,
  \dodoi{10.1089/ast.2012.0936}

\bibitem[{Zahnle(1986)}]{zahnle1986photochemistry}
Zahnle, K.~J. 1986, Journal of Geophysical Research, 91, 2819,
  \dodoi{10.1029/jd091id02p02819}

\end{thebibliography}
\bibliographystyle{aasjournal}



\end{document}